\documentclass[english]{article}
\usepackage{authblk}
\usepackage[T1]{fontenc}
\usepackage{hyperref}
\usepackage{amsmath,amssymb}
\usepackage{url}
\usepackage{appendix}
\usepackage[latin9]{inputenc}
\usepackage{geometry}
\usepackage[english]{babel}
\usepackage{color}
\usepackage{caption}
\usepackage{bm}
\title{\bf Cortical credit assignment by\\Hebbian, neuromodulatory and inhibitory plasticity}
\usepackage{natbib}
\usepackage{graphicx}

\date{}

\begin{document}

\author[1]{Johnatan Aljadeff}
\author[2,3]{James D'amour}
\author[2,3]{Rachel E. Field}
\author[2,3,4]{Robert C. Froemke}
\author[1]{Claudia Clopath}
\affil[1]{\small{Department of Bioengineering, Imperial College London, London, SW7 2AZ, UK}}
\affil[2]{\small{Skirball Institute for Biomolecular Medicine, Neuroscience Institute, Department of Otolaryngology, Department of Neuroscience and Physiology, New York University School of Medicine, New York, NY, 10016, USA.}}
\affil[3]{\small{Center for Neural Science, New York University, New York, NY, 10003, USA.}}
\affil[4]{\small{Faculty Scholar, Howard Hughes Medical Institute, Chevy Chase, MA, 20815, USA}}
\maketitle
\vspace{-0.9cm}
\abstract{The cortex learns to make associations between stimuli and spiking activity which supports behaviour. It does this by adjusting synaptic weights. The complexity of these transformations implies that synapses have to change without access to the full error information, a problem typically referred to as `credit-assignment'. However, it remains unknown how the cortex solves this problem. We propose that a combination of plasticity rules, 1) Hebbian, 2) acetylcholine-dependent and 3) noradrenaline-dependent excitatory plasticity, together with 4) inhibitory plasticity restoring E/I balance, effectively solves the credit assignment problem. We derive conditions under-which a neuron model can learn a number of associations approaching its theoretical capacity. We confirm our predictions regarding acetylcholine-dependent and inhibitory plasticity by reanalysing experimental data. Our work suggests that detailed cortical E/I balance reduces the dimensionality of the problem of associating inputs with outputs, thereby allowing imperfect `supervision' by neuromodulatory systems to guide learning effectively.}

\part*{Introduction}

The cortex learns to accurately process high-dimensional sensory information to support behaviour. Perceptual decisions (e.g., ``I am looking at a dog'') rely on extracting complex features from sensory scenes along a hierarchy \citep{Tanaka1996, Serre2007}. Neuronal responses within a single stage are correlated with many stimulus features, and are highly heterogeneous. Credit assignment refers to the process whereby animals' behavioural performance results from  synaptic changes at every stage of the hierarchy. It remains poorly understood how the cortex solves this problem. There is no consensus whether the basic credit assignment computation is done at the dendritic, synaptic, neuronal or network scale \citep{Friedrich2011, Bourdoukan2015, Richards2018, Sacramento2018}. Previous studies showed that algorithms such as error backpropagation, used to solve difficult classification tasks \citep{Krizhevsky2012}, can be approximated by local learning rules, where weight modifications only depend on that synapse's efficacy, pre- and postsynaptic activity  \citep{Richards2018, Sacramento2018}. These learning rules assume that top-down feedback endows the network with the full error information to adjust its synapses. 

We argue that the complex relationship between single neuron activity and the behavioural outcome makes it probable that not all synapses have access to the explicit value of the error. We focus on extracting error information from neuromodulatory signals available in real cortical circuits. Our model neuron is localised to a single stage of processing and isolated from recurrent feedback, i.e., a perceptron \citep{Rosenblatt1961,Novikoff1962, Mezard1986,Kanter1987,Gardner1988,Amit1989b,Nadal1990,Seung1992,Brunel2004}. When using the full error information, synaptic weights are usually found using the Delta-rule \citep{Rosenblatt1961,Novikoff1962}, where weights change proportionally to the negative error times the input. This ensures that errors decrease over time.

Here, we propose that excitatory synapses follow Hebbian plasticity gated by neuromodulatory signals \citep{Seol2007, Pawlak2010}. We focus on acetylcholine (ACh) and noradrenaline (NE) which were shown to be important for perceptual learning \citep{McGaughy2002,Heath2006,Doucette2007,Froemke2007,Froemke2013,Hangya2015, Martins2015}. Learning in our model is indirectly and imperfectly supervised. Stimulus pairing with a neuromodulator is statistically, but not explicitly,  dependent on the target, in accordance with the specificity of ACh and NE effects on cortical neurons \citep{McCormick1985,Everitt1997,Weinberger2004, Zaborszky2013, Hangya2015, Totah2018, Lacefield2019}, responding to paired and non-paired stimuli \citep{Kilgard1998, Froemke2007,Froemke2013, Chen2012, Martins2015}. Furthermore, model definitions are consistent with task-specific variables known to correlate with ACh and NE levels, including changes to behavioural performance \citep{McGaughy2002,Yu2005,Dayan2006,Hangya2015,Faraji2018}. In addition to their role in gating plasticity, ACh and NE were shown to affect neuronal responses through their excitability \citep{Xiang1998,Rasmusson2000,Joshi2016}, and by activating neurons in superficial and deep cortical layers, either through disinhibition or directly \citep{Froemke2007,Letzkus2011,Letzkus2015,Chen2012,Brombas2014,McCormick1985,McCormick1986,Hedrick2015}. In our model, the post-synaptic neuron sums excitatory and inhibitory currents, and current contribtions of ACh and NE. Following a growing body of experimental \citep{Maffei2008, Dorrn2010, Xue2014, Damour2015, Adesnik2017} and theoretical literature \citep{Vogels2011, Luz2012}, our model also includes inhibitory plasticity which acts to restore excitatory (E)/inhibitory (I) balance.

To understand the combined effect of these mechanisms, we derive a reduced mathematical description of the learning dynamics. Using this theory, we obtain conditions that guarantee robust learning of a number of associations close to the theoretical capacity. One condition is that excitation and inhibition follow detailed balance (a stronger requirement than the global balance in \cite{Rubin2017}). 

Our theory leads to predictions which we test here by reanalysing previously published experiments \citep{Damour2015,Froemke2013}.  Confirming our predictions, inhibitory plasticity restores a detailed balance characterised by a linear relationship between E and I that favours excitation. Furthermore, models that yield realistic synaptic weight distributions (as in \cite{Brunel2016}) predict that ACh-dependent depression outweighs potentiation in a stimulus-dependent fashion. This counterintuitive prediction is consistent with reanalysis of data in \cite{Froemke2013}, and with additional experiments showing, for example, that applying a muscarinic cholinergic receptor agonist preferentially enables associative LTD in cortical slices \citep{Seol2007}; and that the classical hippocampal LTP protocol (pre$\to$post) results in LTD when applying ACh (\citep{Brzosko2017}, see additional references therein).

Finally, using the model we identify the specific roles played by each plasticity mechanism in learning the task. ACh makes small weight changes when currents are close to, but on the wrong side of, the spiking threshold-- it effectively minimises  expected uncertainty. NE makes large modifications when currents associated with a spike output are far below threshold-- it effectively minimises unexpected uncertainty \citep{Yu2005}. Hebbian plasticity ensures that responses are robust to noise by driving currents away from the spiking threshold. Tight E/I balance reduces the dimensionality of the learning problem, so that partial error information contained in neuromodulatory signaling suffices for convergence.

\part*{Results}

\section*{Plasticity mechanisms together approximate the Delta-rule}

\begin{figure}[ht!]
	\begin{center}
		\includegraphics[scale=0.43]{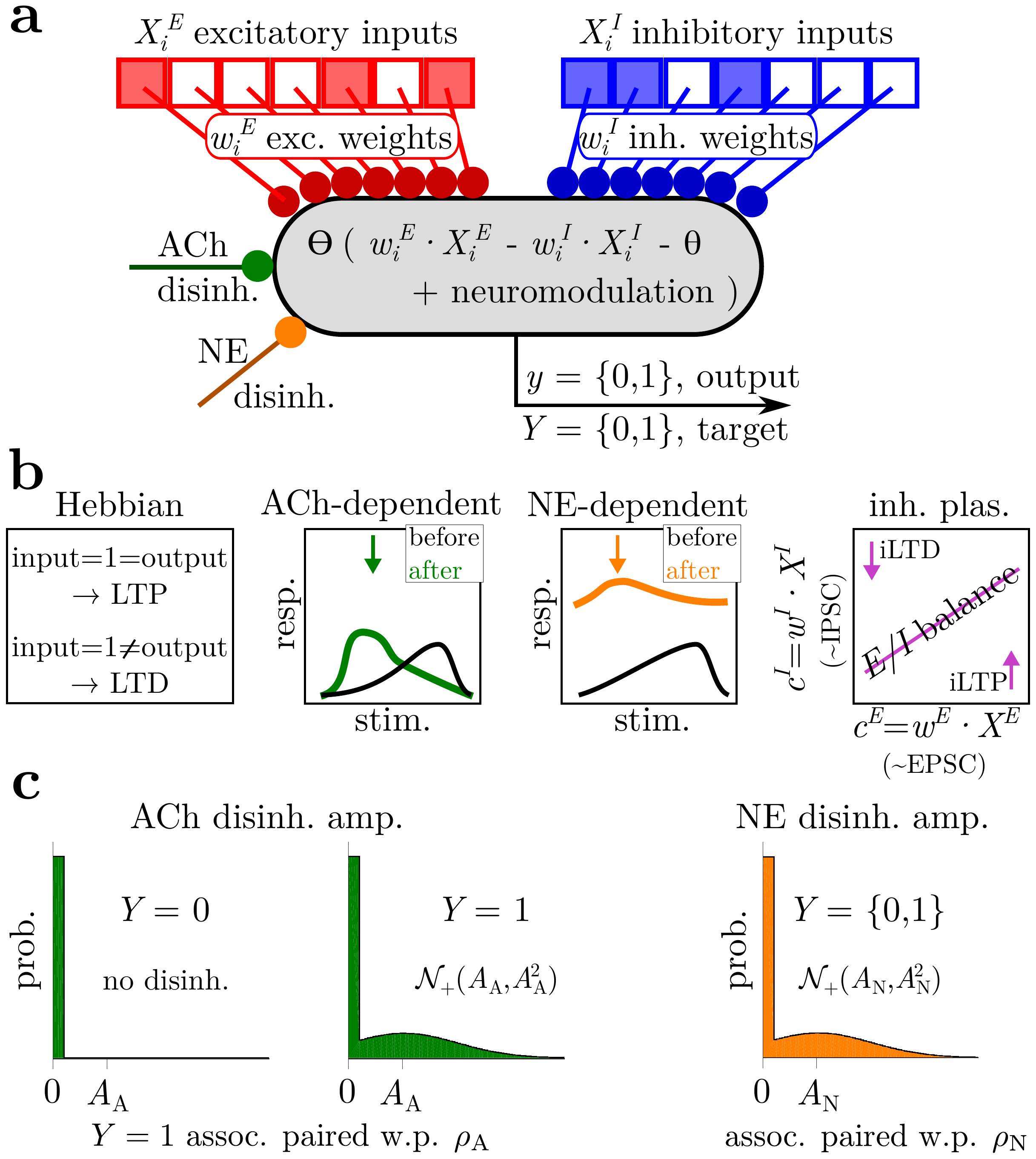}
	\end{center}
	\caption{{\bf Schematic illustration of model components.} ({\bf a}) Neuromodulated binary perceptron model: $N_E$ excitatory and $N_I$ inhibitory random binary inputs weighted by the weight vectors $w^E$ and $w^I$ and added to disinhibitory currents mediated by ACh and NE. The resulting net current is compared with a threshold $\theta$ to produce a spike, or fail to do so. The weights must be modified such that for each of the $P$ input--output associations, the output $y$ is equal to the target output $Y$. ({\bf b}) Schematic of modelled plasticity mechanisms. Hebbian: active E synapses potentiate/depress if postsynaptic neuron is active/inactive and the presynaptic input is active. ACh dependent: E synapses corresponding to active inputs paired with ACh potentiate while those corresponding to inactive paired inputs depress. NE dependent: all E synapses potentiate. I plasticity: drives neuron toward E/I balance. ({\bf c}) Schematic of ACh and NE disinhibitory amplitude statistics included in the model. ACh is paired only with $Y=1$ associations, with probability $\rho_{\rm A}$ per trial. Upon pairing, the amplitude follows a Gaussian distribution with mean and standard deviation $A_{\rm A}$, rectified to exclude negative currents. NE is paired with all associations with probability $\rho_{\rm N}$ per trial. Upon pairing, the amplitude for each association follows a (rectified) Gaussian distribution with mean and standard deviation $A_{\rm N}$.}\label{fig:schematic}
\end{figure}

How do neurons learn to associate inputs with outputs? To address this problem, we study a model neuron receiving high-dimensional excitatory (E) and inhibitory (I) inputs weighted by E, I synapses. The neuron's output (denoted $y$) is a spike if the net current is above a threshold $\theta$. We model one neuron within a large network that processes information hierarchically, so the ``task'' is to respond to each input vector with the target output. E, I inputs and the target output are drawn randomly from a distribution, and denoted $X_{\mu}^E$, $X_{\mu}^I$, $Y_{\mu}$. The input-output associations are indexed by $\mu$, running from $1,\dots,P$, the number of associations in the task (see Supp. Mat. \ref{sec:appdef}). Synaptic plasticity at E, I synapses ($w^E$, $w^I$) leads to learning (see Fig. \ref{fig:schematic}a for illustration of the E, I perceptron model).

The Delta-rule is a plasticity rule guaranteed to find synaptic weights correctly associating inputs with outputs, provided that a solution exists. It decreases the net current in response to an input if the neuron produced an erroneous spike output; and it increases the net current for an erroneous no-spike. This rule therefore uses the error explicitly ($y-Y_{\mu}$, including its sign) to generate a supervisor signal used to update each synaptic weight.

We propose an alternative learning rule composed of four mechanisms acting in parallel. Here, plasticity does not depend explicitly on the target output $Y_{\mu}$, nor does it depend on the error. For two types of plasticity, Hebbian E and I plasticity, weight changes only depend on pre- and postsynaptic activity (i.e., unsupervised learning). In the remaining two, ACh- and NE-dependent E plasticity, weight changes depend also on neuromodulatory ``gates''. Importantly, the probability for opening the ACh gate depends on the target $Y_{\mu}$. In addition to its effect on plasticity, pairing stimuli with neuromodulators (i.e., open ACh and NE gates) also leads to disinhibitory currents. ACh- and NE-dependent currents are added to the E, I currents, and are denoted $M_{\rm A}$, $M_{\rm N}$ (see green, orange lines in Fig. \ref{fig:schematic}a). Thus, neuromodulation implements indirect and imperfect supervision, and the question is whether that suffices for credit assignment. The neuron's output is mathematically expressed as
\begin{equation}
y = \Theta \left( \sum_{i=1}^{N_E} w^E_i X_{\mu,i}^E - \sum_{i=1}^{N_I} w^I_i X_{\mu,i}^I  + M_{\rm A} + M_{\rm N} - \theta \right).\label{eq:EIperceptron}
\end{equation}
Here, $i$ indexes the input components and the weights (i.e., it runs from 1 to $N_E$/$N_I$ for E/I inputs/weights). The Heaviside step-function $\Theta$ is equal to 1 if the argument is positive, and to 0 otherwise.

We note that the probabilistic nature of ACh- and NE-dependent disinhibition is reminiscent of the perturbation step of `trial-and-error' learning, while neuromodulatory gating of plasticity is reminiscent of the evaluation step (see mathematical analysis of the model's learning dynamics, Supp. Mat. \ref{sec:L}, \ref{sec:N}).

\paragraph*{Cholinergic gating of plasticity.} Nucleus Basalis (NB) is the primary mid-brain source of cortical cholinergic projections, carrying information on the behavioural relevance of sensory stimuli \citep{Mesulam1983,Butt1997,Kilgard1998}. Pairing NB stimulation with a stimulus leads to stimulus specific potentiation \citep{Bakin1996, Kilgard1998, Froemke2007, Froemke2013, Chen2012}. It also leads to depression of responses to other stimuli, especially when the paired/unpaired stimuli are dissimilar \citep{Froemke2007, Froemke2013}. 

For simplicity, inputs in our model are binary (i.e., stimulus components are equal to 0 or 1). ACh-dependent LTD/LTP following presentation of the stimulus $X^E_{\mu}$ is,
\begin{eqnarray}
(\Delta w^E_i)_{\rm A, LTP} &  = & + \,\eta_{\rm A} \,\alpha_{\rm A}\,  (y-\bar{f}) \, X^E_{\mu,i} \nonumber \\
(\Delta w^E_i)_{\rm A, LTD} &  = & - \,\eta_{\rm A} \,\alpha_{\rm A}\,  (y-\bar{f}) \, \beta_{\rm A}\,  (1-X^E_{\mu,i}).\label{eq:DwA}
\end{eqnarray}
The gate indicating whether $X^E_{\mu}$ was paired with ACh (denoted $\eta_{\rm A}$), the learning rate (denoted $\alpha_{\rm A}$), and dependence on postsynaptic activity are identical for LTP/LTD. The output $y$ is compared to $\bar{f}$ (reference spike probability for neuromodulatory plasticity). The factor $X^E_{\mu,i}$ ensure that synapses encoding the paired input are potentiated. Conversely, the factor $1-X^E_{\mu,i}$ ensures that synapses not encoding the paired input are depressed. Therefore, changes to the total excitatory response to an input depend on the overlap between that input and the paired input $X^E_\mu$. The relative amount of ACh-dependent LTD/LTP is controlled by a scaling factor (denoted $\beta_{\rm A}$).

Ref. \citep{Weinberger2004} provided some of the strongest experimental evidence for induction of {\em stimulus specific} plasticity by pairing NB stimulation with auditory stimuli.  NB-cortex projections are anatomically and functionally specific \citep{Zaborszky2013,Hangya2015}.  Furthermore, properties of cortical neurons and the organisation of cortical circuits can affect how ACh modulates responses of principal cortical neurons \citep{McCormick1985,Letzkus2011,Wozny2011,Chen2012}, so stimulus specificity may also arise postsynaptically to NB. Finally, recent work shows that stimulus representations in NB are plastic, which allows ACh-dependent disinhibition to contribute to successful auditory fear learning even when the sound and the aversive stimulus are separated in time \citep{Guo2019}. Together these indicate a significant degree of statistical dependence between sensory stimuli and ACh activation, as well as specificity of the effects ACh neuronal activity in NB carry on neurons in the cortex. In the model, an input is paired with ACh with probability $\rho_{\rm A}$ if the target is a spike and with probability 0 for no-spike targets. Upon pairing with ACh, the postsynaptic neuron is disinhibited by a current $M_{\rm A}$ (see Supp. Mat. \ref{sec:appdef}, Fig. \ref{fig:schematic}c). 

\paragraph*{Noradrenalin gating of plasticity.} Locus Coeruleus (LC) is the primary brainstem source of cortical noradrenalin projections. Potentiation following LC stimulation is strong compared to NB stimulation \citep{Martins2015,Froemke2007}, while the specificity is initially less precise \citep{Sara2009, Edeline2011, Eldar2013,Martins2015}.  NE-dependent excitatory plasticity is then modelled as,
\begin{equation}
(\Delta w^E_i)_{\rm N} = \eta_{\rm N}\, \alpha_{\rm N} \, (y-\bar{f}) \, X^E_{\mu,i},\label{eq:DwN}
\end{equation}
where $\eta_{\rm N}$ is the NE gate, and $\alpha_{\rm N}$ is the NE learning rate. NE-dependent LTP is not stimulus specific in our model. Moreover, possible differences between LTP of responses to paired vs. non-paired stimuli (as in \citep{Martins2015}) are negligible for large number of associations, since in each trial stimuli are much more likely to be non-paired than paired. Therefore, $\rho_{\rm N}$ and statistics of the NE disinhibitory current ($M_{\rm N}$) are independent of the target (see Supp. Mat. \ref{sec:appdef}, Fig. \ref{fig:schematic}c).

\paragraph*{Hebbian plasticity.} Hebbian plasticity of E synapses is not gated, and the output $y$ is compared with the reference level $f$ (the average target output, i.e., synaptic changes follow the covariance rule),  
\begin{equation}
(\Delta w^E_i)_{\rm H} = \alpha_{\rm H}\,  (y - f ) \, X^E_{\mu,i},\label{eq:DwH}
\end{equation}
where $\alpha_{\rm H}$ is the Hebbian learning rate.

\paragraph*{Inhibitory plasticity restoring E/I balance.} Inhibitory plasticity can act to restore E/I balance \citep{Maffei2008, Dorrn2010, Xue2014, Damour2015, Adesnik2017}, the characteristics of which (e.g., plasticity timescale, interneuron types involved, and the E/I ratio itself) depend on brain area, developmental stage and species. We define a ``reduced E-I space'' spanned by the E and I currents following presentation of an input $X_\mu$,
\begin{equation}
c^E_\mu = \sum_{i=1}^{N_E} w^E_i X_{\mu,i}^E,\qquad c^I_\mu = \sum_{i=1}^{N_I} w^I_i X_{\mu,i}^I. \label{eq:2dv}
\end{equation}
Note that the encoding of high-dimensional inputs by synaptic weights is now described by a position in the 2-dimensional space $(c^E_\mu,c^I_\mu)$. The spiking threshold is the line $c^I_\mu = c^E_\mu - \theta$. We can think of $c^E_\mu$, $c^I_\mu$ as the EPSC, IPSC following a stimulus, allowing us to later on test model predictions.

Inhibitory plasticity in our model drives each input towards an ``E/I balance line'' (with slope $a$ and offset $b$ that are free parameters), 
\begin{equation}
c^I_\mu = a  c^E_\mu + b. \label{eq:DetailedBalance}
\end{equation}
Changes to inhibitory synapses are then,
\begin{equation}
\Delta w^I_i = \alpha_{\rm I} \,  X^I_{\mu,i}   \, \left[ \left(a  c^E_\mu + b\right) - c^I_\mu \right], \label{eq:DwI} 
\end{equation}
where $\alpha_{\rm I}$ is the inhibitory learning rate. The third factor drives the system towards the E/I balance line (Eqs. (\ref{eq:DwA}, \ref{eq:DwN}, \ref{eq:DwH}, \ref{eq:DwI}) hold as long as weights are between 0 and an upper bound. See details in Supp. Mat. \ref{sec:appdef}, Tables \ref{tab:redmodoptim}, \ref{tab:redmodoptimtilt}).

\subsection*{A reduced model of learning}
Analysing whether the combination of plasticity mechanisms can successfully learn is intractable in the full model. Instead, we derive a reduced model. We determine 1) whether one association can be learned and 2) whether learning one association can overcome interference due to learning other associations. Additionally, we use the reduced model to optimise model parameters in order to learn as many associations as possible. 

Learning dynamics in the reduced E-I space are quantified by averaging the effects of plasticity over all mechanisms. The effective weight update $\mathcal{L}({\bf c}_\mu,Y_\mu)$ describes changes to the net current following a ``test'' stimulus (i.e., the stimulus indexed by $\mu$, whose learning is under question). $\mathcal{L}$ depends on the stimulus' position in the reduced E-I space, ${\bf c}_\mu = (c^E_\mu,c^I_\mu)$, and on the associated target (the subscript $\mu$ is dropped whenever possible, see derivation of $\mathcal{L}$ in Supp. Mat.  \ref{sec:L}). Learning one association ($P=1$) is guaranteed if the net current ($c^E - c^I$) increases for erroneous no-spike outputs and decreases for erroneous spike outputs (i.e., plasticity and disinhibition should, on average, extract the sign of the error and change the synaptic weights similarly to the Delta-rule). 

\begin{figure}
	\begin{center}
		\includegraphics[scale=0.43]{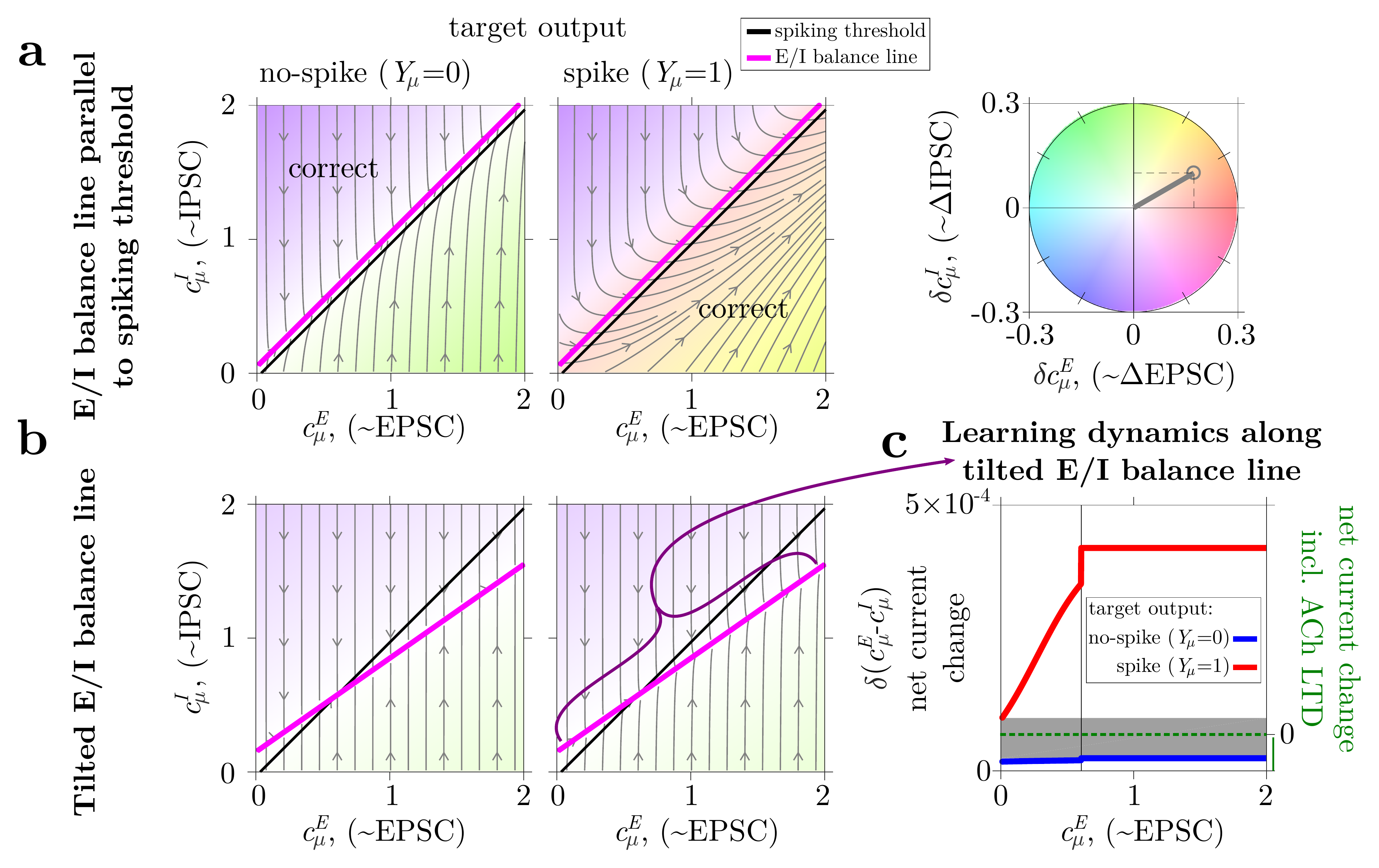}
	\end{center}
	\caption{{\bf Learning in reduced weight space.} ({\bf a}) We plot the angle of the learning induced movement in the 2D space spanned by the excitatory and inhibitory currents in response to presentation of the input $X_\mu$ [i.e., $c^E_\mu-c^I_\mu$, see Eq. (\ref{eq:2dv})]. Points in this space represent a specific E, I input ($\mu$) projected on the E, I weights. If the target output is no-spike (left), correct classification occurs for points above the black line representing the spiking threshold, and the opposite for inputs for which the correct response is a spike (right). Here the spiking threshold (black) is parallel to the E/I balance line (magenta), and inhibitory plasticity is weak. Successful learning  through plasticity drives points to the appropriate region in the reduced weight space (below the black line for $Y_\mu=1$ and above it for $Y_\mu=0$). Mathematically, the learning dynamics are described by a two dimensional vector field ($\delta c_{\mu}^E,\delta c_{\mu}^I$), graphically represented by the gray field lines and by the colour. The rightmost panel shows how colour represents both the direction of weight updates and their magnitude, with white indicating $\delta c_{\mu}^E = \delta c_{\mu}^I = 0$. As an example, the orange color within the gray circle represents $(\delta c_{\mu}^E,\delta c_{\mu}^I) = (0.17,0.1)$ per input presentation. ({\bf b}) Same as (a) for the case where the E/I balance line is tilted with respect to the spiking threshold, and intersects with it, with strong inhibitory plasticity. Here, inhibitory plasticity drives inputs projected to this space towards the magenta lines, where the neuromodulation-dependent plasticity and disinhibition ``sort'' each input to the correct side of the spiking threshold based on the target output. The strong inhibitory plasticity masks the output dependence of the learning drive, so the left and right panels are very similar to one another. ({\bf c}) Learning dynamics of inputs as a function of the excitatory current, restricted to the E/I balance line, when the target output is 0 (blue) and 1 (red). Here inhibitory plasticity is 0 by definition, making the output dependence of plasticity is apparent. Only the effect of the ``test'' stimulus is taken into account, so the net effect of learning is always positive (i.e., $\delta(c^E_{\mu}-c^I_{\mu})>0$). The depression induced by ACh-dependent plasticity when ``background'' stimuli are presented is chosen such that neutral learning dynamics ($\delta(c^E_{\mu}-c^I_{\mu})=0$) are in the gray-shaded area. Doing so ensures that net excitation increases for inputs corresponding to a target spike, and decreases for no-spike target, so that a large number of associations can be learned.}\label{fig:reduced}
\end{figure}

Movements of the E, I currents ($c^E,c^I$) relative to one another during the learning process is governed partially by inhibitory plasticity [Eqs. (\ref{eq:DetailedBalance}, \ref{eq:DwI})]. We therefore studied two qualitatively different definitions of E/I balance. First, the E/I balance line and spiking threshold (Fig. \ref{fig:reduced}a, magenta and black, respectively) are parallel. Here inhibitory plasticity drives the system towards an E/I ratio of $\approx$1, and the E/I balance line is wholly in the no-spike region. Learning one association in this case requires weak inhibitory plasticity. If this plasticity is strong it would drive E, I currents to the balance line, that is to the no-spike region, independent of the target. We will show later that using this definition gives a model that cannot learn a large number of associations.  Second, we define the E/I balance line to intersect with the spiking threshold (Fig. \ref{fig:reduced}b). Owing to this intersection, E, I currents can be balanced and lead to a spike or no-spike response. Here, strong inhibitory plasticity drives currents fast to the E/I balance line, independent of the target. The initial movement towards the E/I balance line is followed by movement along it (Fig. \ref{fig:reduced}c),  which serves to sort the currents to the correct side of the spiking threshold. Note that the net change due to presentation of one input (i.e., the test stimulus) is always positive, even for a no-spike target (Fig. \ref{fig:reduced}c, blue). This does not take into account LTD of inputs not paired with ACh, since the effective weight update $\mathcal{L}$ is computed based on a test input. 

When learning more than one association ($P>1$), changes to the net current following a stimulus must overcome changes due to the presentation of the remaining stimuli. We therefore compute the interference, denoted $\mathcal{N}(\mathbf{c})$, which describes changes to the input current in response to the test stimulus $\mu$, following plasticity due to presentation of a ``background'' stimulus $\nu$ (see derivation of $\mathcal{N}$ in Supp. Mat.  \ref{sec:N}). We can now write the conditions,
\begin{align}
\mathcal{L}(\mathbf{c},Y=1) + (P-1) \mathcal{N}(\mathbf{c}) > 0 & \qquad \text{if} \qquad c^E - c^I - \theta < 0,  \nonumber \\
\mathcal{L}(\mathbf{c},Y=0) + (P-1) \mathcal{N}(\mathbf{c}) < 0 & \qquad \text{if} \qquad c^E - c^I - \theta > 0. \label{eq:learncond1}
\end{align}
For two associations ($P=2$), these conditions were converted into an objective function for optimisation, denoted $\mathcal{C}$, quantifying the characteristic effective weight update {\em relative} to the interference (i.e., $\mathcal{L}/\mathcal{N}$; see Supp. Mat. \ref{sec:numoptim}). $\mathcal{C}$ is an order-of-magnitude estimate for the maximal number of associations that can be learned simultaneously. We use it to show that the conditions [Eq. (\ref{eq:learncond1})] can be satisfied everywhere in the reduced E-I space, including along the E/I balance line. More specifically, understanding how learning of one association affects another (by estimating the interference $\mathcal{N}$) is essential because the effect of the test stimulus alone (described by $\mathcal{L}$) is always positive (Fig. \ref{fig:reduced}c). However, ACh-dependent LTD of responses to background stimuli (included in $\mathcal{N}$) ensures for an appropriate choice of parameters that the net current increases when the target is a spike, and decreases for a no-spike target. We also use the function $\mathcal{C}$ to systematically search for the best parameters. For $P>2$, ensuring that these conditions are satisfied is intractable, since the location of each pair of (test, background) associations is different. Thus we use the results from reduced model with $P=2$ and carry out numerical simulations of the full model with $P\gg 1$.

\section*{Model performance in high dimensions}

Next we investigated the algorithm's ability to learn synaptic weights classifying a set of $P$ high-dimensional inputs. Using $\mathcal{C}$, we optimised parameters under the two qualitatively different definitions of inhibitory plasticity described above. When the E/I balance line is parallel to the spiking threshold (magenta, black lines in Fig. \ref{fig:full}a, as in Fig. \ref{fig:reduced}a), our estimate for the maximal number of associations that can be learned in this scenario is small, $\mathcal{C}\approx 10$ (Table \ref{tab:redmodoptim}). This is far below the theoretical capacity \citep{Rubin2017}. This discrepancy is understood intuitively by plotting the learning dynamics of all inputs in the reduced E-I space (Fig. \ref{fig:full}a). Inputs associated with a target spike ($Y_\mu=1$) are initially correctly classified. Inhibitory plasticity causes these associations to drift towards the E/I balance line which is wholly in the no-spike region, leading to a misclassification. This version of the model fails because inhibitory plasticity makes learning of all inputs corresponding to an output spike unstable.

\subsection*{Successful models follow ``tilted'' and detailed E/I balance}\label{sec:tilt}

We conclude that to learn a large number of associations, the E/I balance line must be tilted relative to the spiking threshold, so that the two intersect in the reduced E-I space (Fig. \ref{fig:full}c). We henceforth restrict ourselves to models where the slope of the E/I balance line is less than unity (i.e., $a<1$). A scenario where the two lines intersect due to $a>1$ is not plausible, since that would require responses to inputs paired with ACh to be depressed rather than potentiated (see Supp. Mat. \ref{sec:appdef} for further discussion). Furthermore, we focus on models where inhibitory plasticity is strong relative to the other mechanisms in our model. This means that E, I follow detailed balance (i.e., all associations lie on the E/I balance line, Fig. \ref{fig:full}c), and that fine-tuning of inhibitory plasticity is no longer necessary. From a theoretical standpoint, assuming that inhibitory plasticity is strong allows us to analyse the combined effect of neuromodulatory and Hebbian plasticity separately from the effect of inhibitory plasticity (Fig. \ref{fig:reduced}c), since they operate on different timescales (see Supp. Mat. \ref{sec:LNb}). 

\subsection*{The neuromodulatory pairing rates $\bm{\rho_{\rm A},\,\rho_{\rm N}}$}\label{sec:FindRates}

The effect of ACh and NE plasticity depends on the total number of associations to be learned, $P$, because stimulus specificity differs between ACh and NE. Therefore, some model parameters controlling ACh and NE plasticity must depend on $P$. We choose to make the pairing rates ($\rho_{\rm A},\,\rho_{\rm N}$) depend on $P$ as it is known that the activity of NB and LC neurons projecting to cortex depends on the task difficulty, novelty and the animal's performance and behavioural state \citep{Butt1997,Hironaka2001,Pudovkina2001,Reimer2016}.

We evaluate the learning performance for each set of associations on a grid in the 2-dimensional parameter space $\rho_{\rm A}-\rho_{\rm N}$ (other parameters remaining constant), and find large variability of the best pairing rates (Fig. \ref{fig:rho}). We therefore allow $\rho_{\rm A},\,\rho_{\rm N}$ to adjust dynamically, using limited information on the average performance. Specifically, changes to $\rho_{\rm A}$ are proportional to the difference between errors occurring when the target is spike/no-spike. Changes to $\rho_{\rm N}$ are proportional to the difference between target and current average output (see Supp. Mat. \ref{sec:rhot}). This leads to a large increase in the classification performance for arbitrary instances of the learning task and the stability of the solution weight vectors. Overcoming input correlations arising from finite size effects further suggests that ``online'' adjustment of neuromodulation pairing rates may allow our algorithm to successfully learn associations that include correlations in the limit $N\to \infty$. Therefore our algorithm may be flexible enough to learn associations even when the environment and the stimulus statistics change.

\subsection*{Classification performance is close to theoretical capacity}\label{sec:compare}

We now compare our algorithm's classification performance to the theoretical capacity. \cite{Rubin2017} computed the theoretical capacity for segregated E, I inputs and binary output. Imposing E/I balance reduces the capacity (especially when $N_E > N_I$), but leads to an increased robustness to output noise. Importantly, the global E/I balance in \cite{Rubin2017} is a weaker requirement than the detailed balance, necessary here for learning using our algorithm. Intuitively, balance is {\em detailed} if the E/I balance condition [Eq. (\ref{eq:DetailedBalance})] holds for each of the $P$ inputs, while global balance involves a single sum of the weights. We obtain an upper bound for the theoretical capacity of a perceptron under {\em detailed} E/I balance, and explain why statistical physics techniques developed in \citep{Gardner1988,Amit1989b,Nadal1990,Brunel2004,Rubin2017} cannot be directly applied here (see Supp. Mat. \ref{sec:BalanceCapacity}).

\vspace{0.5cm}

\begin{figure}[ht!]
	\begin{center}
		\includegraphics[scale=0.43]{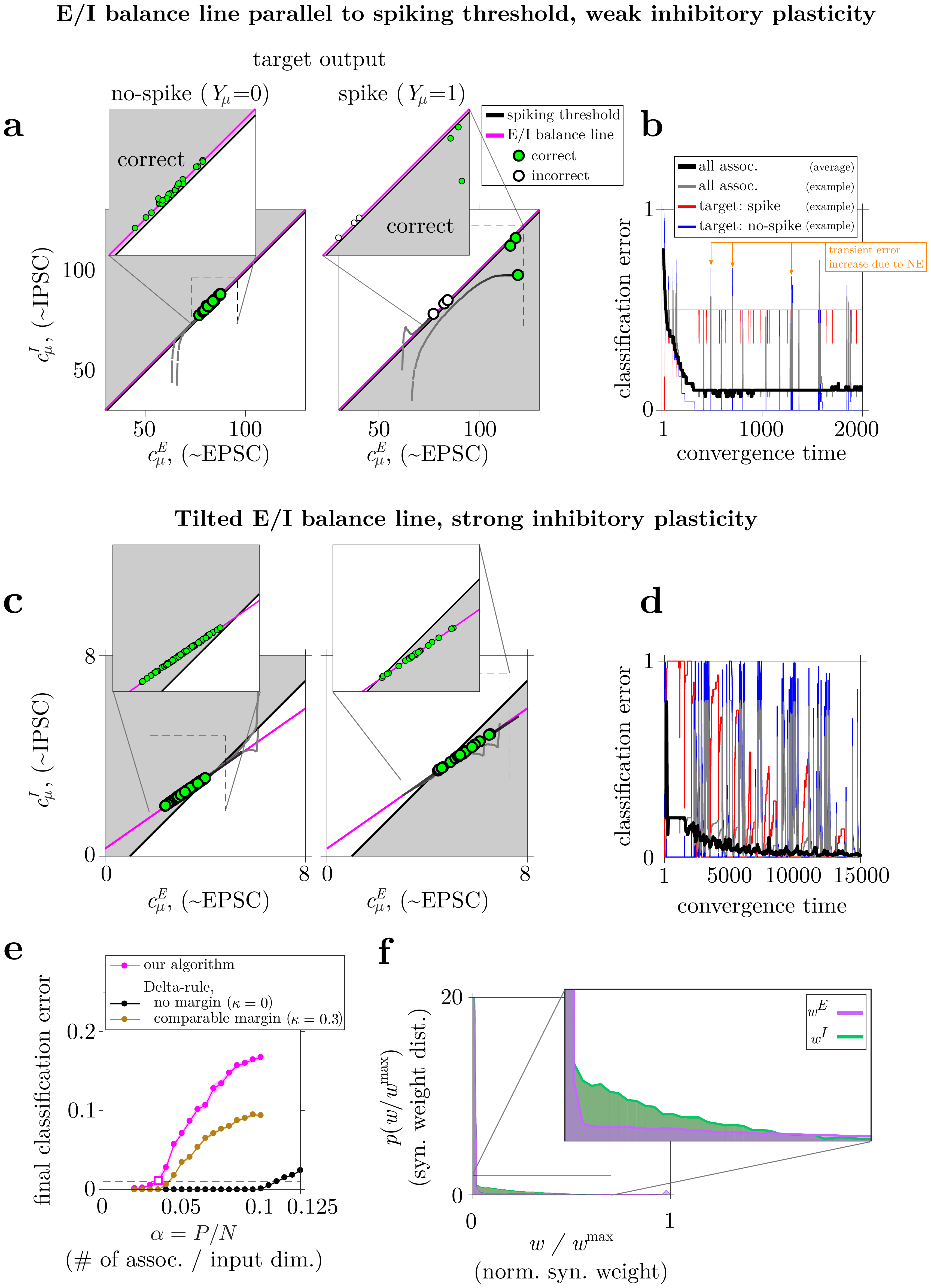}\\
		Caption on next page.
	\end{center}
\end{figure}

	\newpage
	\captionof{figure}{{\bf Learning in full model: only ``tilted'' E/I balance leads to successful learning of a larger number of associations.} ({\bf a}) When the spiking threshold (black) is parallel to the E/I balance line (magenta), the model fails to learn $P=30$ associations, despite the fact that inhibitory plasticity is weak. Initial conditions were chosen such that inputs corresponding to an output spike ($Y_\mu=1$) are correctly classified. Gray lines indicate the trajectory over 2000 presentations of each stimulus for example associations, and circles indicate position of all associations at the end of learning (green/white-- correct/incorrect classification). Gray shading indicates regions in the reduced E-I space where classification is correct. ({\bf b}) Evolution of misclassification rate during one example simulation for $Y_\mu=0$ associations (blue), $Y_\mu=1$ associations (red), over all associations (gray); and the total error averaged over 48 simulations (black). Transient increases in error rates are due to pairing with NE that potentiate excitation in response to all inputs (orange arrows). Hebbian and neuromodulatory plasticity acts to keep $Y_\mu=1$ associations above the spiking threshold, but the inhibitory plasticity drives them below it after a transient, leading to a high misclassification rate of these inputs (red). ({\bf c}) Same as (a) for a model where the E/I balance line is tilted relative to the spiking threshold, and strong inhibitory plasticity. Here the size of the learning problem is $P=140$. Associations initially move fast towards balance line, then are learned by moving above/below the spiking threshold based on pairing with neuromodulatory inputs that carry information on the correct response. ({\bf d}) Same as (b) for model with tilted E/I balance line. The error is close to 0 except transiently upon pairing with NE. ({\bf e}) Classification error as a function the ``load'' $\alpha$, the number of associations $P$ divided by the input dimension $N=N_E+N_I$ for 48 simulations for each value of $P$. We compare our algorithm to the Delta-rule including the detailed balance constraint, and no margin for error (black), or margin matched to the result of our simulations (brown). When the margin for error is matched (see Fig. \ref{fig:Kappa}), our algorithm correctly classifies a number of associations close to the Delta-rule. Square symbol corresponds to the examples in panels (c,d). ({\bf f}) Resulting synaptic excitatory  (purple) and inhibitory (green) synaptic weight distributions, for $P=140$. Many synapses saturate to the lower bound, yet few excitatory and no inhibitory synapses saturate to the upper bound, implying that our model is consistent with data showing unimodal synaptic weight distributions (input dimensions used: $N_E = 3200,\,N_I= 800$).\label{fig:full}
}

\vspace{0.5cm}

We compute the capacity under detailed balance numerically, by alternating Delta-rule weight updates and updates to I weights [Eqs. (\ref{eq:DetailedBalance}, \ref{eq:DwI})]. We used two values of the margin parameter $\kappa$. A non-zero margin means that only solutions that are robust to fluctuations of amplitude $\kappa$ of the net current are admitted. Setting $\kappa = 0$, we find that the capacity is approximately three times larger than the maximal number of inputs classified using our algorithm (Fig. \ref{fig:full}e, black). This however is not a fair comparison between the two algorithms, because the average distance of the net current from the spiking threshold is larger using our algorithm than using the Delta-rule (compare magenta and black lines in  Fig. \ref{fig:Kappa}). We therefore found $\kappa$ for which the average distance of the net current from the spiking threshold matches between our algorithm and the Delta-rule (Fig. \ref{fig:Kappa}, magenta and brown lines). For this value of $\kappa$, we find that our algorithm learns to classify a number of patterns equal to 80-90\% of the theoretical capacity (Fig. \ref{fig:full}e, brown).

To quantify the amount of information given to the neuron by the neuromodulatory systems, we compute the mutual information between the target output and the neuron's response, including possible ACh- and NE-dependent currents, by averaging over the statistics of E, I currents and the output statistics. Information is degraded by the small probability of a neuromodulatory gate being open, and by the probability that neuromodulatory currents will fail to lead to a correct output. We compare this to the error information provided by the Delta-rule by assuming that the neuron's response (for the purpose of learning) is always correct. Our algorithm (for the parameters used in Fig. \ref{fig:full}c) provides $0.014$ bits of ``error information'' in each stimulus presentation, whereas the Delta-rule provides $0.722$ bits (a 50-fold difference).

\begin{figure}[ht!]
	\begin{center}
		\includegraphics[scale=0.43]{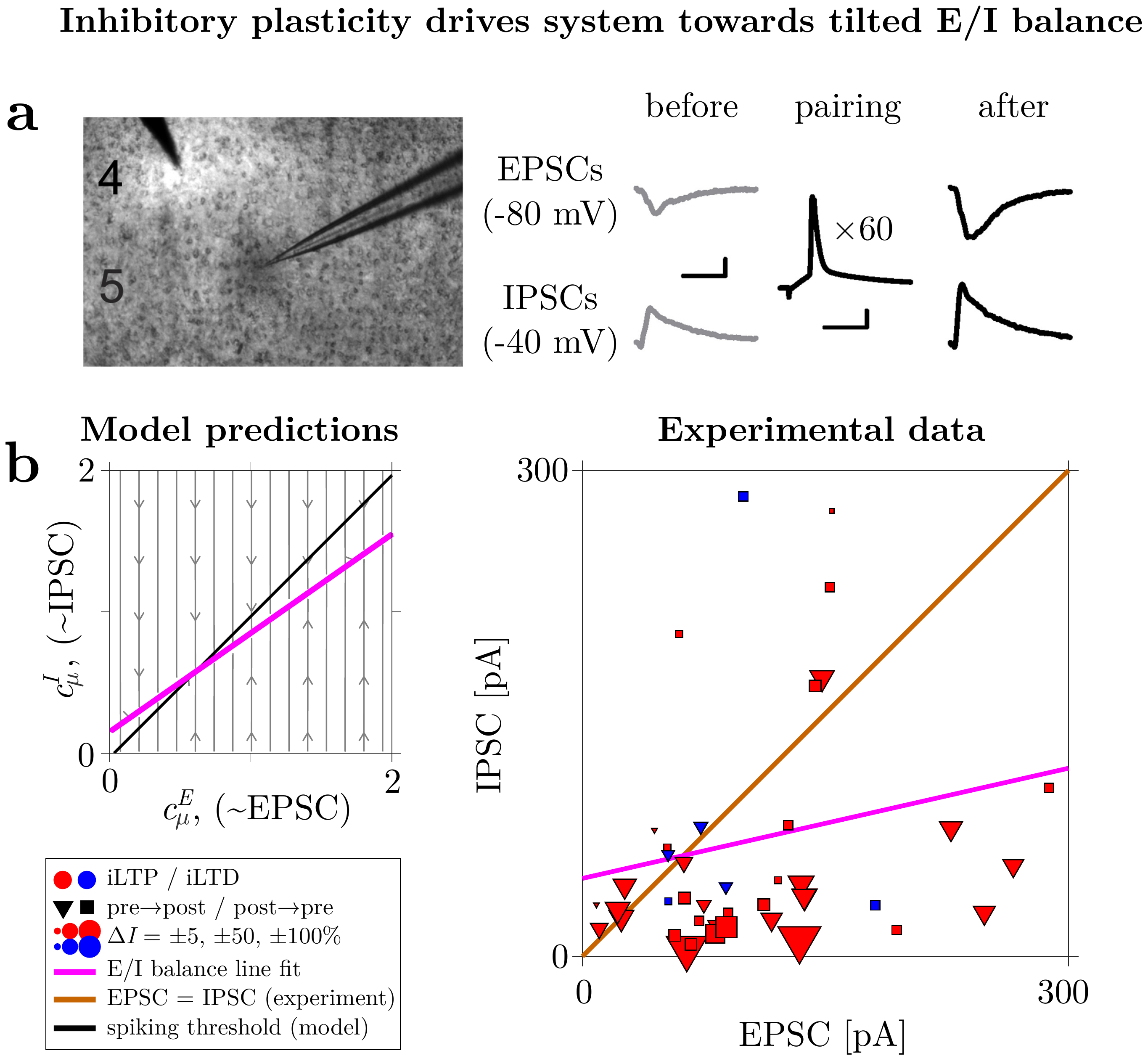}
	\end{center}
	\caption{{\bf Validation of theoretical predictions: inhibitory plasticity measured {\em in vitro} \citep{Damour2015} drives towards a tilted E/I balance line.} ({\bf a}, left) Image of whole-cell recording from layer 5 pyramidal neuron of mouse auditory cortex in brain slice. ({\bf a}, right) Experimental design. Extracellular stimulation evoked EPSCs and IPSCs monitored at $-80$ mV and $-40$ mV, respectively, in voltage-clamp before and after pairing. Scale: $10$ ms, $100$ pA. During pairing, recordings were switched to current-clamp to allow postsynaptic cells to fire single action potentials paired with single extracellular shocks (60 pairings, 0.1-0.2 Hz). Scale: 5 ms, 20 mV. Panel (a) is reproduced from \citep{Damour2015}. ({\bf b}) Data triplets [EPSCs, IPSCs, inhibitory plasticity, denoted ($C_E,\,C_I,\,\Delta I$)] were pooled over pre$\to$post (triangles) and post$\to$pre (squares) protocols. Symbol size corresponds to the magnitude of LTD/LTP. We fit $\Delta I$ as a function of $C_E,\,C_I$ [Eq. (\ref{eq:DwIexp2})], which then allows us to plot the E/I balance line (magenta) by setting $\Delta I=0$. All LTD (blue, $\Delta I<0$) and all LTP (red, $\Delta I>0$) data points were given the same weight, with the relative weight between LTD and LTP chosen such that they have the same total weight. We excluded outliers (defined as points with EPSC or IPSC $>300$pA). The fit gives parameters consistent with the requirements for successful learning extracted from the model, shown in the inset (see Table \ref{tab:Ifit} for quantitative fit results, and Fig. \ref{fig:tilt2} for fit results without excluding outliers).}\label{fig:experiment}
\end{figure}

\subsection*{Predictions of the model: ``Tilted'' E/I balance}

Our theoretical analysis shows that only a tilted E/I balance line that intersects with the spiking threshold supports learning a large number of associations. We tested our prediction by reanalysing \citep{Damour2015}. In \citep{Damour2015}, we used a Spike-Timing-Dependent-Plasticity (STDP) protocol to measure changes to inhibitory synaptic weights in auditory cortical slices. To assess the dependence of plasticity on the relationship between E and I currents, these currents were measured before pairing by holding the neuron at the E, I reversal potentials (Fig. \ref{fig:experiment}a). We examined the relations between the initial synaptic strengths quantified by the EPSCs and IPSCs, and inhibitory modifications,denoted ($C_E,\,C_I,\,\Delta I$). We pool over pre$\to$post and post$\to$pre protocols, since differences between these had a relatively weak effect on plasticity, and since our model does not include a timing component. We fit the parameters of the equation
\begin{equation}
\Delta I =  \frac{\alpha_{\rm I}}{C_I} \left( A C_E + B- C_I \right),\label{eq:DwIexp2}
\end{equation}
equivalent to Eq. (\ref{eq:DwI}), with an inhibitory learning rate that is inversely proportional to the IPSC. Setting $B=0$ gives a linear relationship between $\Delta I$ and the E/I ratio, as in \citep{Damour2015}. Our theory predicts that the learning rate will be positive $\alpha_{\rm I}>0$, ensuring that inhibitory plasticity is driving the system towards E/I balance. Furthermore, to ensure that the balance line is tilted and intersects with the spiking threshold, we predict that $1>A>0$, and $B>0$ (details of the numerical fit are included in Supp. Mat. \ref{sec:Ifit}). All three parameters agree with our theoretical prediction. Moreover, including the parameter $B$ in the fit reduces the root-mean-squared-error (RMSE) of the model's fit to the data by 6\% (Table \ref{tab:Ifit}). The non-zero value of $B$ is important because it implies that inhibitory plasticity depends on the EPSC and IPSC the not only through the E/I ratio. Finally, we have slope $A<1$ (Fig. \ref{fig:experiment}b, magenta, similar to the model shown in the inset), as was also found in a number of additional studies, on the timescale of minutes-hours \citep{Maffei2008, Xue2014, Adesnik2017}, and days-weeks \citep{Dorrn2010}.

E/I balance is also characterised by the offset $b$ [Eq. (\ref{eq:DwI})]. To relate this quantity to data, we study its effect on the resulting synaptic weight distributions after learning (Fig. \ref{fig:offset}). When $b$ is large and the intersection of the E/I balance line and the spiking threshold is far from the point EPSC=IPSC=0, the synaptic weight distribution becomes bimodal (Fig. \ref{fig:offset}a,b).  This is in contrast to experimentally measured synaptic weight distributions that do not exhibit bimodalities \citep{Markram1997,Sjostrom2001, Song2005,Lefort2009,Brunel2016}. We thus conclude that a model consistent with experimental data regarding synaptic weight distributions, should have a small offset of the E/I balance line relative to the spiking threshold $\theta$.

\subsection*{Predictions of the model: Depression bias of ACh-dependent plasticity}

\begin{figure}[ht!]
	\begin{center}
		\includegraphics[scale=0.43]{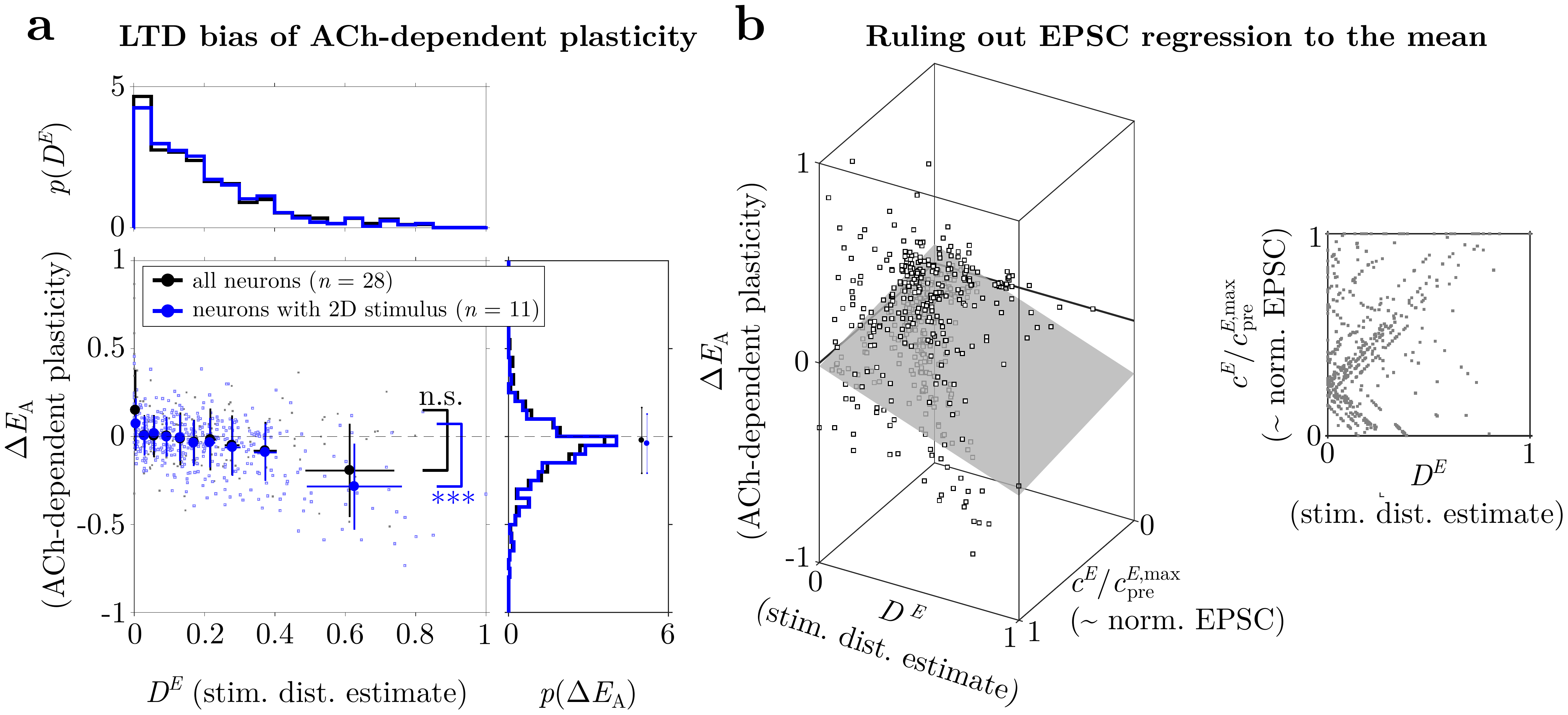}
	\end{center}
	\caption{{\bf Validation of theoretical predictions: ACh-dependent plasticity measured {\em in vivo} \citep{Froemke2013} is LTD biased.} We computed $D^E$, the absolute difference of normalised EPSCs between the stimulus paired with ACh stimulation and every other stimulus the animal was probed with. We assume that $D^E$ is anti-correlated with the overlap of these stimuli in the space of synaptic inputs. Our model predicts that LTD of excitatory responses to stimuli distant from the one paired with ACh should be larger than LTP of excitation in response to near stimuli. (a) We plot the normalised change in EPSC $(\Delta E_{\rm A})$ as a function of $D^E$, separately for cells recorded when animals were probed with a two-dimensional stimulus set (frequency$\times$intensity, blue) and a one-dimensional stimulus set (frequency/intensity, gray).  Large black/blue circles show the average dependence of plasticity on stimulus distance, in 10 equally sized bins. We tested our prediction that LTD for large distances should be stronger than LTP for smaller distances by comparing the data points in the first and last bin (in terms of $D^E$). Restricting the analysis to experiments with a two-dimensional stimulus space gives a significant effect ($P<10^{-5}$, two-sided {\em t}-test, $n=34$ points in each group). Taking all data points into account, the effect is consistent with our prediction but not statistically significant ($P = 0.096$, two-sample {\em t}-test, $n=54$ points in each group). The distribution of $D^E$ (top inset) has less values close to 0 for 2-dimensional vs. 1-dimensional stimuli, suggesting that inputs are more dissimilar for higher stimulus dimensions. Note that the stronger depression we predicted and found in the data does not contradict the fact that no significant change in excitation was found when averaging over all stimuli and all neurons (right inset, circles show the distribution average), since this is likely a result that depends on the stimulus set. ({\bf b}) To rule out the possibility that large LTD of distant stimuli is a regression to the mean effect whereby large EPSCs are preferentially depressed, we linearly regressed $(\Delta E_{\rm A})$ against both $D^E$ and the normalised EPSC before pairing $c^E/c^{E,\max}_{\rm pre}$, giving $\Delta E_{\rm A} = 0.095 - 0.360 \, D^E - 0.112 \, c^E/c^{E,\max}_{\rm pre}$. We therefore conclude that stronger LTD of distant stimuli compared to LTP of nearby stimuli is better explained by the distance ($\sim$overlap) between stimuli (similar to our model) than by regression to the mean. The correlation between $D^E$ and $c^E/c^{E,\max}_{\rm pre}$ is small ($R^2=0.111$, right), despite that $c^E/c^{E,\max}_{\rm pre}$ appears explicitly in the definition of $D^E$.}\label{fig:experiment2}
\end{figure}

Consider the change to the  excitatory current in response to an input $X^E$ (denoted $\Delta E_{\rm A}$) as a function of the overlap $r^E = X^E\cdot X^E_{\rm pair}$ (where $X^E_{\rm pair}$ is the input paired with ACh stimulation). Our definition of ACh-dependent plasticity [Eq. (\ref{eq:DwA})] implies that currents undergo LTP when $X^E$ and $X^E_{\rm paired}$ are highly overlapping ($r^E\approx  1$), and depressed for nearly orthogonal stimuli. The parameter $\beta_{\rm A}$ controls the strength of ACh-dependent LTD relative to LTP. When the offset of the E/I balance line is small, $\beta_{\rm A}$ leading to successful learning should be significantly larger than 1 (favouring LTD, see Fig. \ref{fig:offset}c, Table \ref{tab:redmodoptimtilt}). Intuitively, a bias towards LTD for small $b$ is expected because the intersection of the spiking threshold and the E/I balance line determines a ``set-point'' where currents concentrate in the reduced E-I space (see Fig. \ref{fig:full}, Fig. \ref{fig:offset2}). When this set point is close to EPSC=IPSC=0, many weights are saturated to 0 and cannot depress further. Saturation in turn necessitates larger $\beta_{\rm A}$, since LTD of unpaired synapses is less effective than implied by $\beta_{\rm A}$.

We test these predictions by analysing \citep{Froemke2013}. In \cite{Froemke2013}, we paired nucleus basalis stimulation with an auditory tone at a specific intensity and frequency while measuring EPSCs in an auditory cortical neuron.  There was no direct experimental measurement of the high-dimensional input ($\sim X^E$), but rather the EPSC ($\sim c^E = w^E \cdot X^E$). We assume that current similarity is correlated with stimulus overlap, justified by the unimodal tuning of EPSCs. In agreement with our prediction, ACh-dependent LTD for large absolute differences in currents ($D^E = |\hat{c}^E-\hat{c}^E_{\rm paired}| \approx 1$) is stronger than LTP for similar inputs ($D^E \approx 0$). This effect becomes stronger when the analysis is restricted to neurons probed with a two-dimensional stimulus (compare blue and black points, Fig. \ref{fig:experiment2}a), likely caused by smaller stimulus overlap when both tone frequency and intensity are varied. Note that the overall neutrality of changes in excitation reported in \citep{Froemke2013} does not contradict our finding of bias towards LTD, since the synthetic stimulus ensemble in the experiment makes it difficult to draw conclusions based on averaging $\Delta E_{\rm A}$ over all stimuli and neurons. We performed an additional analysis ruling out the possibility that the LTD bias is caused by regression of EPSCs to the mean (see Supp. Mat \ref{sec:LTDbias}, Fig. \ref{fig:experiment2}b). 

We conclude that the ACh-dependent plasticity, as reported in \citep{Froemke2013} and in additional experiments \citep{Seol2007,Brzosko2017}, supports our theoretical prediction for a bias towards LTD. We therefore think that this form of plasticity is well-poised to play a role (as ascribed in our model) in extracting the error information necessary for credit assignment.

\part*{Discussion}

Credit assignment is a multifaceted problem. Temporal credit assignment \citep{Gutig2006,Gutig2016,Friedrich2011} concerns adjusting neuronal activity unfolding over time based on outcomes  that may be delayed by hundreds of milliseconds. Credit assignment has also spatial aspects, since error information must propagate from the sensory periphery through the processing hierarchy \citep{Friedrich2011, Richards2018, Sacramento2018,Lansdell2019}. Previous studies typically rely on gradient-based weight-updates, i.e., they assume that the full error information exists in the network, at least in one stage of the hierarchy and at one point in time. Less attention has been given to the question of the origin of the error signal, especially in tasks involving classification of high-dimensional inputs. Here we propose a model inspired by experimental literature that learns such tasks, where multiple plasticity mechanisms together translate the output into useful weight modifications.

\paragraph*{Roles of plasticity mechanisms.}
Learning in our model is based on three algorithmic ingredients: stochastic perturbation of neuronal responses; dimensionality reduction of the input space; which then allows the response (given the task) to be evaluated. We identify how each ingredient is implemented by mechanisms in our model.

{\bf Inhibitory plasticity.} We showed that indirect supervision by neuromodulatory signals--as defined in our model--is by itself insufficient to learn a large number of associations (Fig. \ref{fig:full}a,b). The learning rule cannot keep inputs in the solution region due to interference between associations. This problem is solved by restricting inputs to a line in the reduced E-I space, enforced in the model by inhibitory plasticity driving the system towards detailed E/I balance with specific properties. Thus, inhibitory plasticity reduces the problem dimensionality (see \citep{Bourdoukan2015,Deneve2016}). The reduced input dimension allows partial error information (equal to a small fraction, $\approx 1/50$, of that provided by the Delta-rule) to suffice for learning.

{\bf Neuromodulator-dependent plasticity and disinhibition.} Perturbation of neuronal responses is implemented by randomly pairing neuromodulation-dependent disinhibitory currents with inputs. Neuromodulation also gates plasticity that, in turn, depends on postsynaptic activity, thereby creating a link between the perturbation and evaluation of the response. Our model is defined to match stimulus specificity of neuromodulatory-dependent plasticity. ACh causes potentiation/depression of synapses encoding/not-encoding the paired inputs; while NE potentiates responses to all inputs \citep{Froemke2007,Froemke2013,Martins2015}. For the learning algorithm to work, ACh should lead to smaller disinhibitory currents and smaller synaptic changes than NE, consistent with the types of learning thought to be mediated by these neuromodulators. 

\cite{Yu2005} proposed that ACh signals expected uncertainties, while NE signals unexpected uncertainties. We interpret expected uncertainty as inputs encoded close to the spiking threshold, where small amounts of noise lead to erroneous threshold crossings. Small ACh currents that depend on the target response ``sort'' inputs to the correct side of the threshold. In contrast, inputs associated with a target spike that are far below threshold are unexpected. Here, crossing threshold requires large NE mediated currents. Note however that the weak stimulus specificity of NE implies that responses to inputs corresponding to a no-spike target are also potentiated, leading to transient increases of erroneous spike responses (Fig. \ref{fig:full}d). Thus, we interpret the strong and non-specific effect of NE as causing ``world model'' updates.

{\bf Hebbian plasticity.} Hebbian plasticity in our model (Eq. \ref{eq:DwH}) leads to LTP of E synapses if the postsynaptic neuron spikes following an input, and to LTD if there is no spike. Therefore, if an input is correctly classified, Hebbian plasticity drives the net current away from the spiking threshold, making the encoding more robust to noise (Fig. \ref{fig:alphaHebb}).

\paragraph*{Other proposed mechanisms of extracting error information.}

In \cite{Bouvier2018}, we proposed that cerebellar learning at parallel fibre to Purkinje cell synapses is implemented by plasticity rules equivalent to {\em stochastic} gradient descent. That modeling work asked how such a biologically plausible rule could support learning of an analog (``firing rate'') perceptron \citep{Clopath2013}. The cerebellar model does not require two ``flavours'' of perturbation, because graded Purkinje cell responses admit incremental error reduction. The two models nevertheless share the three algorithmic components discussed above. \cite{Whittington2017} also avoided making explicit use of the error in a deep network, by ``clamping'' the output to the correct one in the learning phase, akin to the so called ``target-propagation'' algorithm in machine learning \citep{LeCun1986,Bengio2014} (see Supp. Mat. \ref{sec:supdisc} for further discussion of connections with literature).

The brain learns to produce behaviours by coordinating contributions from neurons in different regions, where there is a highly complex relationship between errors and appropriate synaptic weight modifications. By focusing on a single layer, our analysis shows that multiple plasticity mechanisms acting in concert are needed to effectively extract information necessary for trial-and-error learning. From an experimental point of view, our study highlights the importance of investigating how plasticity depends on neuromodulation. Theoretically, our work provides a possibility of perturbation-based learning algorithms could be implemented in the brain. Extending our results to deep or recurrent architectures will help us understand the computational roles played by extensive yet specific cortical projections from the mid-brain and from higher areas; as well as the roles of heterogeneous plasticity rules in learning.

\subsection*{Acknowledgements}
We thank N. Brunel and G. Bouvier for comments on the manuscript.

\subsection*{Author contributions}
Research design and model development: J.A. and C.C.; Theoretical model analysis and simulations: J.A., with contributions from C.C.; Data analysis: J.A., with contributions from C.C. and R.C.F.; Contribution of electrophysiological data: J.D., R.E.F., and R.C.F.; Writing: J.A. and C.C., with contributions from R.C.F.

\subsection*{Code and data availability}
Computer code will be made available upon publication. Data will be made available upon request.

\newpage
\setcounter{section}{0}
\renewcommand\thesection{S\arabic{section}}
\numberwithin{equation}{section}

\part*{Supplementary Material}

\section{Mathematical definitions of model}\label{sec:appdef}

\paragraph{E, I perceptron with neuromodulation.}
We study a binary neuron learning random input--output associations
\begin{equation}
y = \Theta \left( \sum_{i=1}^{N_E} w^E_i X_{\mu,i}^E - \sum_{i=1}^{N_I} w^I_i X_{\mu,i}^I  + M_{\rm A} + M_{\rm N} - \theta \right).
\end{equation}
Here, $w^E$ and $w^I$ are excitatory (E) and inhibitory (I) synaptic weight vectors of dimensions $N_E$ and $N_I$, respectively; $X_{\mu,i}^E$ and $X_{\mu,i}^I$ are the E and I components of the $\mu$ input ($\mu=1,\dots,P$, $i=1,\dots,N_E/N_I$); $M_{\rm A}$ and $M_{\rm N}$ are acetylcholine (ACh)  and noradrenaline (NE) neuromodulatory inputs; and $\theta$ is the spiking threshold. The output $y$ is equal to 1 if the argument of the Heaviside step-function $\Theta(\cdot)$ is positive, and to 0 otherwise. The goal of the learning task is to find $w^E$ and $w^I$ such that for each $\mu$ the output $y$ is equal to a target value $Y_\mu$. The input--output statistics (i.e., the probability distributions from which $X_{\mu}^E$,  $X_{\mu}^I$ and $Y_\mu$ are drawn) are parameters describing the learning task. For simplicity we assume that $X_{\mu,i}^E$,  $X_{\mu,i}^I$ and $Y_\mu$ are Bernoulli random numbers equal to 1 with probability $f$ and to 0 with probability $1-f$. 

\paragraph{Plasticity rules and disinhibition.}

\begin{description}
\item[ACh-dependent excitatory plasticity] is defined to be
\begin{equation}
(\Delta w^E_i)_{\rm A} = \eta_{\rm A} \alpha_{\rm A} (y - \bar{f} ) \left[ X^E_{\mu,i} - \beta_{\rm A} \frac{f}{1-f} (1-X^E_{\mu,i})\right].
\end{equation}
Here, $\eta_{\rm A}$ is the ACh gating variable, indicating whether the input $X_{\mu}$ is paired with ACh or not ($\eta_{\rm A}=0,\,1$, respectively); $\alpha_{\rm A}$ is the ACh learning rate; $\bar{f}$ is the reference spike probability for neuromodulatory dependent plasticity, with $\bar{f}\approx 0$, since responses to paired stimuli typically are not depressed even if the postsynaptic neuron does not spike ($y=0$). The parameter $\beta_{\rm A}$ describes the balance between potentiation of responses to paired inputs and depression of responses to unpaired inputs.  The factor $f/(1-f)$ ensures that for $\beta_{\rm A}=1$ potentiation and depression are balanced when the effects of ACh-dependent plasticity are averaged over the input-output statistics. An input $X^E_\mu$ is paired with ACh (i.e., $\eta_{\rm A}=1$) with probability $1>\rho_{\rm A}>0$ if the target $Y_\mu = 1$ and with probability 0 if $Y_\mu = 0$. When an input $X^E_\mu$ is paired with ACh, the postsynaptic neuron is disinhibited by an amplitude $M_{\rm A}$ drawn from a rectified Gaussian distribution with mean and standard-deviation $A_{\rm A}$. 

\item[NE-dependent excitatory plasticity] is defined to be
\begin{equation}
(\Delta w^E_i)_{\rm N} = \eta_{\rm N} \alpha_{\rm N} (y - \bar{f} ) X^E_{\mu,i},
\end{equation}
where $\eta_{\rm N}$ is the NE gating variable, and $\alpha_{\rm N}$ is the NE learning rate. Here the probability that an input is paired with NE is $1>\rho_{\rm N}>0$, {\em independent} of the target output $Y_\mu$. When an input $X^E_\mu$ is paired with NE, the postsynaptic neuron is disinhibited by an amplitude $M_{\rm N}$ drawn from a rectified Gaussian distribution with mean and standard-deviation $A_{\rm N}$. 

\item[Hebbian excitatory plasticity] is defined to be 
\begin{equation}
(\Delta w^E_i)_{\rm H} = \alpha_{\rm H} (y - f ) X^E_{\mu,i},
\end{equation}
where $\alpha_{\rm H}$ is the Hebbian learning rate. 

\item[Inhibitory plasticity] is defined to be
\begin{equation}
\Delta w^I_i = \alpha_{\rm I} \left[ \left(a  c^E_\mu + b\right) - c^I_\mu \right] X^I_{\mu,i} ,
\end{equation}
where $\alpha_{\rm I}$ is the inhibitory plasticity learning rate and the E, I currents are defined in Eq. (\ref{eq:2dv}). We also consider slight modifications of this rule where changes to inhibitory synapses occur only if the postsynaptic neuron spikes (i.e., $\Delta w^I_i\to y  \Delta w^I_i$, in accordance with the results in \citep{Damour2015}), or to depend nonlinearly on the synaptic weight (i.e., $\Delta w^I_i\to g(w^I_i)  \Delta w^I_i$). We find that in versions of the model that can robustly learn the task there is negligible difference between these alternative definitions.
\end{description}

\paragraph{Model simulations.}
Inputs are presented sequentially in random order. Every time a stimulus is presented the gating and disinhibitory variables $\eta_{\rm A},\,\eta_{\rm N}\,M_{\rm A},\,M_{\rm N}$ are drawn randomly as described above, and used to compute the output $y$ and the resulting plasticity. The synaptic weights are then updated according to
\begin{eqnarray}
w^E_i(t+1) & = & \min\left\{\left[w^E_i(t) + (\Delta w^E_i)_{\rm A} +(\Delta w^E_i)_{\rm N} + (\Delta w^E_i)_{\rm H}\right]_+ ,w^E_{\max}\right\}, \nonumber \\
w^I_i(t+1) & = & \min\left\{\left[w^I_i(t) + \Delta w^I_i\right]_+ ,w^I_{\max}\right\},
\end{eqnarray}
where $w^E_i(t)$, $w^E_i(t+1)$ denote the E weights before and after the input is presented (and similarly for inhibition). 

\paragraph{Slope of E/I balance line.} We note that in principle the E/I balance line and spiking threshold could intersect if $a>1$. This however is inconsistent with evidence from the neuromodulatory-dependent plasticity literature. To see this, consider the values of $c^E$ {\em along the balance line} that lead to a spike/no-spike response. If $a<1$, $c^E$ corresponding to a spike is greater than that corresponding to no-spike, while the opposite is true if $a>1$. So $a>1$ implies that excitatory weights encoding a specific sensory input should {\em depress} upon pairing of that input with neuromodulatory stimulation, in clear contradiction to evidence from the ACh and NE systems \citep{Froemke2007,Froemke2013,Martins2015}.

\section{The effective weight update $\mathcal{L}$}\label{sec:L}

The probability that ACh are paired with an input $X$ depends on the correct output $Y$, so $\mathcal{L}$ is computed separately for $Y=0,1$. Recall that the disinhibitory currents $M_{\rm A},\,M_{\rm N}$ are drawn from truncated Gaussian distributions taking the form
\begin{eqnarray}
p\left(M_{\rm A}\right)& =& \frac{1}{\sqrt{2\pi A_{\rm A}^2}}e^{-\frac{1}{2}\left(\frac{M_{\rm A}}{A_{\rm A}}-1\right)^2}\Theta\left(M_{\rm A}\right)+\delta\left(M_{\rm A}\right)\tilde{p}(0,A_{\rm A}),\\
\tilde{p}(d,A) & =&  \frac{1}{2}{\rm erfc}\left(-\frac{d+A}{\sqrt{2}A}\right),\nonumber
\end{eqnarray}
and similarly for $M_{\rm N}$, with $A_{\rm A}\to A_{\rm N}$. If a disinhibitory current $M$ is present and drawn from the probability distribution $p(M)$ with parameter $A$, using the shorthand notation $d({\bf c})=c^E-c^I-\theta$, we can write the probability of spike $q(d({\bf c}),A)$ (i.e., the probability that the net current $d({\bf c}) + M$ is positive),
\begin{eqnarray}
q(d({\bf c}),A) %& = & \begin{cases}1,& d({\bf c})\ge 0\\
		  %          		\int_{-d({\bf c})}^{\infty} p(M)dM, & d({\bf c})<0 \end{cases},\\
		  & = & \begin{cases}1,& d({\bf c})\ge 0\\
							\tilde{p}(d({\bf c}),A), & d({\bf c})<0 \end{cases}					
\end{eqnarray}  

We begin with the case where $Y=0$. When the target output is 0, by assumption there is no ACh modulation, so the probability of an output spike ($y=1$) given the currents is $q(d({\bf c}),A_{\rm N})$. Average changes to excitatory weights are,
\begin{eqnarray}
\Delta w^E_i & = & X^E_i \langle \alpha_{\rm H} (y-f) + \alpha_{\rm N}\eta_{\rm N}  (y-\bar{f})\rangle, \nonumber\\
             & = & X^E_i \left[ \alpha_{\rm H} ((1-\rho_{\rm N})\Theta(d({\bf c})) + \rho_{\rm N}q(d({\bf c}),A_{\rm N})-f) + \alpha_{\rm N}\rho_{\rm N}  (q(d({\bf c}),A_{\rm N})-\bar{f})\right],\label{eq:DwEY0}
\end{eqnarray}
The change in total excitatory current is,
\begin{equation}
\frac{\Delta c^E}{Nf} = \alpha_{\rm H} (\Theta(d({\bf c}))-f) + \rho_{\rm N}\left[ q(d({\bf c}),A_{\rm N})\left(\alpha_{\rm H}+\alpha_{\rm N}\right) - \alpha_{\rm H}\Theta(d({\bf c})) - \alpha_{\rm N}\bar{f}\right].\label{eq:DvEY0}
\end{equation}
Similarly, the change in inhibitory current is,
\begin{equation}
\Delta c^I = Nf \alpha_{\rm I} r^{EI}, \label{eq:DvIY0}
\end{equation}
giving the effective weight update, 
\begin{equation}
\frac{1}{Nf}\mathcal{L}\left({\bf c},0\right) = \alpha_{\rm H} (\Theta(d({\bf c}))-f) + \rho_{\rm N}\left[ q(d({\bf c}),A_{\rm N})\left(\alpha_{\rm H}+\alpha_{\rm N}\right) - \alpha_{\rm H}\Theta(d({\bf c})) - \alpha_{\rm N}\bar{f} \right] - \alpha_{\rm I} r^{EI}.\label{eq:LY0}
\end{equation}
Note that $\mathcal{L}$ depends on $c^E$ and $c^I$ not only through their difference, since $r^{EI} = a c^E - c^I + b$ (and in general $a\neq 1$).

We now move on to study the case $Y=1$. When the target output is 1, an output spike can be due to the net current being above threshold ($d({\bf c})>\theta$) or a the net current plus either an ACh or NE disinhibitory current leading to a threshold crossing. We neglect the possibility that on a given trial an input is paired with {\em both} ACh and NE. The average changes to the excitatory current is now,
\begin{eqnarray}
\frac{\Delta c^E}{Nf} & = & \langle \alpha_{\rm H} (y-f) + \alpha_{\rm A}\eta_{\rm A}(y-\bar{f})+\alpha_{\rm N}\eta_{\rm N}  (y-\bar{f})\rangle, \nonumber\\
             & = & \alpha_{\rm H} \left((1-\rho_{\rm A}-\rho_{\rm N})\Theta(d({\bf c})) + \rho_{\rm A}q(d({\bf c}),A_{\rm A}) +  \rho_{\rm N}q(d({\bf c}),A_{\rm N})-f\right)  \nonumber \\
        & &    + \alpha_{\rm A}\rho_{\rm A}\left(q(d({\bf c}),A_{\rm A}) -\bar{f}\right) + \alpha_{\rm N} \rho_{\rm N}\left(q(d({\bf c}),A_{\rm N})-\bar{f}\right).\label{eq:DvEY1}
\end{eqnarray}
The change in inhibitory current is the same as Eq. (\ref{eq:DvIY0}), so the effective weight update for $Y=1$ is 
\begin{eqnarray}
\frac{1}{Nf}\mathcal{L}\left({\bf c},1\right) & = & \alpha_{\rm H} \left((1-\rho_{\rm A}-\rho_{\rm N})\Theta(d({\bf c})) + \rho_{\rm A}q(d({\bf c}),A_{\rm A}) +  \rho_{\rm N}q(d({\bf c}),A_{\rm N}) -f\right)  \nonumber \\
        & &    + \alpha_{\rm A}\rho_{\rm A}\left(q(d({\bf c}),A_{\rm A}) -\bar{f}\right) + \alpha_{\rm N}\rho_{\rm N}\left(q(d({\bf c}),A_{\rm N})-\bar{f}\right) - \alpha_{\rm I} r^{EI}.\label{eq:LY1}
\end{eqnarray}

The function $\mathcal{L}$ can be thought of as the derivative of the net current $c^E-c^I$ assuming the learning task has a single input--output association. Given a set of model parameters, convergence of learning of a single association is determined by ensuring that evolving the net current from any two dimensional initial condition $(c^E(t=0),c^I(t=0))$ leads to the correct response (i.e., $y = \Theta(d({\bf c})) = Y$).

\section{The interference $\mathcal{N}$}\label{sec:N}
The strategy for computing the interference is similar to that used for computing the effective weight update, with an additional averaging step: previously we averaged over the statistics of the stochastic neuromodulation pairing [e.g., Eq. (\ref{eq:DwEY0})]. Now we also average over the input--output statistics of the ``background'' associations. When doing so we must take into account the average overlap of inputs, which determines by how much learning a background pattern changes the weights that encode the test pattern.

To do this calculation, we replace the two dimensional weight variable used previously with $(c^E,c^I)$ with a six dimensional variable,
\begin{equation}
\mathbf{c} = (\,c_{\mu\nu}^E,\,c_{\mu\bar{\nu}}^E,\,c_{\bar{\mu}\nu}^E,\,c_{\mu\nu}^I,\,c_{\mu\bar{\nu}}^I,\,c_{\bar{\mu}\nu}^I\,).\label{eq:6D}
\end{equation}
Here, $c_{\mu\nu}^E$ denotes the sum of weights corresponding to excitatory input components that are active for $\mu$ {\em and} $\nu$ (i.e., indices $i$ such that $X^E_{\mu,i}=X^E_{\nu,i}=1$); $c_{\mu\bar{\nu}}^I$ denotes the sum of weights corresponding active inhibitory input components for $\mu$ {\em but inactive} for $\nu=2$ (i.e., indices $i$ such that $X^E_{\mu,i}=1-X^E_{\nu,i}=1$); and so on. Formally, using the fact that the $X$'s are binary, we can write,
\begin{equation}
c_{\mu\nu}^E = \sum_{i=1}^{N_E} w^E_i X_{\mu,i}^E X_{\nu,i}^E, \qquad c_{\mu\bar{\nu}}^E = \sum_{i=1}^{N_E} w^E_i X_{\mu,i}^E (1-X_{\nu,i}^E), \qquad
c_{\bar{\mu}\nu}^E = \sum_{i=1}^{N_E} w^E_i (1-X_{\mu,i}^E) X_{\nu,i}^E, \label{eq:6dv}
\end{equation}
and similarly for inhibition.

We rewrite the plasticity outcomes using Eqs. (\ref{eq:DwA}, \ref{eq:DwN}, \ref{eq:DwH}, \ref{eq:DwI}) following presentation of the background pattern $\nu$ in terms of the components of the reduced six-dimensional weight space [Eq. (\ref{eq:6dv})], 
\begin{eqnarray}
\frac{1}{N f^2}\Delta c^E_{\mu\nu} = \frac{1}{N f(1-f)}\Delta c^E_{\bar{\mu}\nu} & = &  \langle (\alpha_{\rm A} \eta_{\rm A} + \alpha_{\rm N} \eta_{\rm N}) (y-\bar{f}) + \alpha_{\rm H} (y-f) \rangle \nonumber \\
  \frac{1}{N f^2}\Delta c^E_{\mu\bar{\nu}} & = &  -\alpha_{\rm A} \beta_{\rm A} \langle \eta_{\rm A} (y-\bar{f}) \rangle \nonumber \\
\frac{1}{N f^2}\Delta c^I_{\mu\nu} = \frac{1}{N f(1-f)}\Delta c^I_{\bar{\mu}\nu} & = &  \alpha_{\rm I} r^{EI}_\nu \nonumber \\
  \Delta c^I_{\mu\bar{\nu}} & =&  0 \label{eq:dvbkg}
\end{eqnarray}
Here, $\langle\cdot\rangle$ denotes averages over the probability of an input being paired with a neuromodulator, and the input--output statistics. To simplify the expressions we separate the computation of the average change into two regions, defined as a function of the synaptic weight variables and the neuromodulation disinhibitory amplitude parameters (Fig. \ref{fig:schema}). When the current is above threshold or close to it, we assume that the effects of neuromodulation on plasticity and on the response (through disinhibition) act separately. Under this assumption we can compute the probability that $y=1$ and the change in synaptic weights by replacing the gating variables with probability of inputs being paired with neuromodulation (i.e., $\eta$'s $\to$ $\rho$'s). This assumption is exact if the weighted inputs are above threshold in the absence of neuromodulation. When the weighted input is far below threshold we assume that the response can only become $y=1$ on a given trial as a result of NE modulation. Now we can compute the average change in bias because the ``source'' of threshold crossing (which is needed to compute the change in synapses) is NE by assumption. As a separator between these regions we use the value $sA_{\rm A}$ ($s=2$). Mathematically, Region 1 is defined to be,
\begin{equation}
\text{Region 1} = \{\mathbf{c}|c^E_\nu-c^I_\nu-\theta + s A_{\rm A} \ge 0 \},
\end{equation}
and Region 2 is everywhere else in the reduced weight space.

\subparagraph*{Region 1-- close to or above threshold.}
We assume that the probability of ACh or NE is the ``cause'' of crossing the firing threshold is proportional to the rate with which inputs are paired with these neuromodulators. The amplitudes are taken into account only in the calculation of the probability of $y=1$. The averages in Eq. (\ref{eq:dvbkg}) are performed by replacing $\eta_{\rm A}\to\rho_{\rm A}$ and $\eta_{\rm N}\to\rho_{\rm N}$, writing the expressions for $\Delta \mathbf{c}$ separately for $y=0,\,1$, and multiplying them by the respective probabilities ($p(y=1|\mathbf{c})=\psi(\mathbf{c})$, $p(y=0|\mathbf{c})=1-\psi(\mathbf{c})$). This probability is,
\begin{equation}
\psi(\mathbf{c}) = \Theta\left(d(c_\nu)\right)+\Theta\left(-d(c_\nu)\right) \left[\rho_{\rm A}f \tilde{p}(d(c_{\nu}),A_{\rm A}) + \rho_{\rm N} \tilde{p}(d(c_{\nu}),A_{\rm N})\right],\label{eq:psi}
\end{equation}
where  $d(c_\nu) = c^E_{\nu}-c^I_{\nu}-\theta$, $c^E_{\nu} = c^E_{\mu\nu}+c^E_{\bar{\mu}\nu}$ and $c^I_{\nu} = c^I_{\mu\nu}+c^I_{\bar{\mu}\nu}$. The interference in Region 1 from the point of view of the test association $\mu$ following presentation of the background association $\nu$ is then, 
\begin{eqnarray}
\frac{1}{Nf^2} \mathcal{N}_1(\mathbf{c}) & = & (\Delta c^E_{\mu\nu} + \Delta c^E_{\mu\bar{\nu}}) + (\Delta c^I_{\mu\nu} + \Delta c^I_{\mu\bar{\nu}}),\nonumber \\
& = & \left[\alpha_{\rm A} \rho_{\rm A}\left(1-\beta_{\rm A}\right) + \alpha_{\rm N} \rho_{\rm N}\right] (\psi(\mathbf{c})-\bar{f}) + \alpha_{\rm H} (\psi(\mathbf{c})-f)- \alpha_{\rm I} r^{EI}_\nu.\label{eq:N1}
\end{eqnarray}

\subparagraph*{Region  2-- far below threshold.}
When the weighted inputs are far below threshold we assume that only NE modulation can cause a threshold crossing. Under this assumption we again compute the average change in synapses when $y =0,\,1$  and jointly the probability of each event. Here we cannot simply average over the target output $Y=0,\,1$ because that influences the probability of neuromodulation which itself modifies the response.

For associations with a target no-spike output ($Y_\nu=0$), changes to the excitatory components of the reduced weight space are, 
\begin{equation}
\frac{1}{N f^2}\Delta c^E_{\mu\nu} = \frac{1}{N f(1-f)}\Delta c^E_{\bar{\mu}\nu} = \rho_{\rm N}( \alpha_{\rm N}  +\alpha_{\rm H}) \tilde{p}(d(c_{\nu}),A_{\rm N})-(\alpha_{\rm N} \rho_{\rm N}\bar{f} + \alpha_{\rm H} f).
\end{equation}
The remaining four components are the same as in Eq. (\ref{eq:dvbkg}). The interference in this case is,
\begin{equation}
\frac{1}{Nf^2} \mathcal{N}_{2,0}(\mathbf{c}) = \rho_{\rm N}( \alpha_{\rm N} +\alpha_{\rm H}) \tilde{p}(d(c_{\nu}),A_{\rm N})-(\alpha_{\rm N} \rho_{\rm N}\bar{f} + \alpha_{\rm H} f + \alpha_{\rm I} r^{EI}_\nu).\label{eq:N20}
\end{equation}
For associations with a target spike output ($Y_\nu=1$), there is direct and indirect NE modulation. By assumption ACh cannot lead to threshold crossing, but it can indirectly contribute to learning. The first two lines of Eq. (\ref{eq:dvbkg}) now read,
\begin{eqnarray}
\frac{1}{N f^2}\Delta c^E_{\mu\nu} = \frac{1}{N f(1-f)}\Delta c^E_{\bar{\mu}\nu} & = &  
\rho_{\rm N}(\alpha_{\rm A} \rho_{\rm A} + \alpha_{\rm N} + \alpha_{\rm H})\tilde{p}(d(c_{\nu}),A_{\rm N}) - \rho_N (\alpha_{\rm A} \rho_{\rm A} + \alpha_{\rm N} ) \bar{f} -\alpha_{\rm H} f, \\
 \frac{1}{N f^2}\Delta c^E_{\mu\bar{\nu}} & =&  -\alpha_{\rm A} \beta_{\rm A} \rho_{\rm A} \left[\rho_{\rm N}\tilde{p}(d(c_{\nu}),A_{\rm N})-\bar{f}\right]\nonumber
 \end{eqnarray}
and the inhibitory components are the same. The interference in this case is,
\begin{eqnarray}
\frac{1}{Nf^2} \mathcal{N}_{2,1}(\mathbf{c})&  =& \rho_{\rm N}(\alpha_{\rm A} \rho_{\rm A}(1-\beta_{\rm A}) + \alpha_{\rm N} + \alpha_{\rm H})\tilde{p}(d(c_{\nu}),A_{\rm N})\nonumber \\
&& - \rho_N (\alpha_{\rm A} \rho_{\rm A} (1-\beta_{\rm A}) + \alpha_{\rm N} ) \bar{f} -\alpha_{\rm H} f - \alpha_{\rm I} r^{EI}_\nu.\label{eq:N21}
\end{eqnarray}
The total interference in Region 2 is simply the average of Eqs. (\ref{eq:N20}, \ref{eq:N21}) over the output statistics, i.e., 
\begin{equation}
\mathcal{N}_2(\mathbf{c}) = (1-f) \mathcal{N}_{20}(\mathbf{c}) + f \mathcal{N}_{21}(\mathbf{c}). 
\end{equation}

\section{Numerical optimisation of model parameters in reduced space}\label{sec:numoptim}

We have derived a reduced model within which we estimate the learning and interference rates of a test and background pattern. We now turn to using the functions $\mathcal{L}$ and $\mathcal{N}$ to numerically optimise the model parameters, as well as obtain an order of magnitude estimate of the algorithm's learning capacity. We define the optimisation's routine objective function $\mathcal{C}$ to be,
\begin{equation}
\frac{1}{\mathcal{C}} = \langle \frac{\mathcal{N}}{\mathcal{L}} \rangle_{X,Y}  =  \frac{f \,\int \frac{\mathcal{N}({\bf c})}{\mathcal{L}({\bf c},Y=1)}\Theta(-d(c_\mu)){\rm d}{\bf c}}{\int \Theta(-d(c_\mu)){\rm d}{\bf c}} +   \frac{(1-f)\,\int \frac{\mathcal{N}({\bf c})}{\mathcal{L}({\bf c},Y=0)}\Theta(d(c_\mu)){\rm d}{\bf c}}{\int \Theta(d(c_\mu)){\rm d}{\bf c}}.\label{eq:optimobj}
\end{equation}
In words, we average the ratio of the effective weight update and the interference rate over the reduced weight-space, conditioned on learning being ``necessary'', i.e., assuming the target response is a spike (no-spike) if the net current is below (above) threshold. The quantity $\mathcal{C}$ which our numerical procedure maximises can be thought of as the largest number of inputs $P$ that the algorithm can learn to classify, because it averages over the input--output statistics the ratio of interference of a single background association ($\mathcal{N}$) to learning a single test association ($\mathcal{L}$). 

We discretised each dimension of the reduced weight-space to $n_b = 10$ values, so Eq. (\ref{eq:optimobj}) was computed over $10^6$ points, and used an optimisation routine built-in MATLAB to find the best set of model parameters. We used parameter bounds indicated in Table \ref{tab:redmodoptim}, and the additional constraints:
\begin{align}
\rho_{\rm N}<\rho_{\rm A}, \qquad & \text{ACh pairing is more frequent than NE pairing,}\nonumber \\
5 A_{\rm A}< A_{\rm N}, \qquad & \text{ACh disinhibition is at least 5 times weaker than NE disinhibition,}\nonumber \\
\alpha_{\rm N}>\alpha_{\rm A}, \qquad & \text{ACh learning rate is smaller than NE learning rate,}\nonumber \\
\alpha_{\rm A}>\alpha_{\rm H}, \qquad & \text{ACh learning rate is larger than Hebbian learning rate.}
\end{align}

\begin{table}
	\begin{center}
		\begin{tabular}{|c|l|c|c|}
		 \hline
 		 Parameter & Meaning & Range & Value \\
   		 \hline\hline
 		 $\alpha_{\rm A}$ & ACh learning rate & [0, 1]& 0.575  \\
 	 	 \hline
		 $\rho_{\rm A}$ & ACh pairing rate & [0, 1]& 0.825 \\
		 \hline
		 $A_{\rm A}$ & ACh disinhibitory amplitude & [0, 2] &0.227 \\
		 \hline
		 $\beta_{\rm A}$ & ACh-dependent LTP/LTD ratio & [0, 5] &0.331  \\
		 \hline
		 $\alpha_{\rm N}$ & NE learning rate  & [0, 1]& 0.772 \\
		 \hline
		 $\rho_{\rm N}$ & NE pairing rate & [0, 1]& 0.012 \\
		 \hline
		 $A_{\rm N}$ & NE disinhibitory amplitude & [0, 2]&1.605  \\
		 \hline
		 $\alpha_{\rm H}$ & Hebbian learning rate & [0, 1] & 0.016 \\
		 \hline
		 $\alpha_{\rm I}$ & Inhibitory learning rate & [0, 1]& 0.638 \\
		 \hline\hline
		 $\mathcal{C}$ & Estimated algorithm capacity & & 10.27 \\
		 \hline  
		\end{tabular} 
	\end{center}
	\caption{Optimised model parameters in the case of parallel spiking threshold and E/I balance lines. Fixed model parameters: $a=1,\,b=0.05$ (balance line slope and offset, respectively), $\bar{f} = 0.01$ (reference spiking level for neuromodulatory plasticity). Task parameters used $f=0.2$, $N_E = 3200,\,N_I = 800$, $\theta = 1$.}		\label{tab:redmodoptim}
\end{table}

\section{Reduced model with detailed E/I balance}\label{sec:LNb}
We showed that for robust learning of a large number of input--output associations, the line defining E/I balance for the purposes of inhibitory plasticity must intersect with the spiking threshold. We made the additional assumption that this balance is {\em tight}: after a short transient, we have $r^{EI}_\mu = a c^E_\mu + b - c^I_\mu = 0$ for all $\mu = 1,\dots,P$ and at all times. This assumption is equivalent to ``separation of timescales'' (between fast inhibitory plasticity and slow excitatory plasticity), and it allows us to further reduce the dimension of the simplified model through which we study the system.

The learning rate $\mathcal{L}$ was previously computed in a two-dimensional space ($c^E_\mu-c^I_\mu$, see Supp. Mat. \ref{sec:L}). Under the additional constraint of E/I balance the space becomes one-dimensional, and terms proportional to $r^{EI}$ drop. Eqs. (\ref{eq:LY0}, \ref{eq:LY1}) become now-- 
\begin{eqnarray}
\frac{1}{Nf}\mathcal{L}\left({\bf c},0\right) & = & \alpha_{\rm H} (\Theta(d({\bf c}))-f) + \rho_{\rm N}\left[ q(d({\bf c}),\beta_{\rm N}A_{\rm N})\left(\alpha_{\rm H}+\alpha_{\rm N}\right) - \alpha_{\rm H}\Theta(d({\bf c})) - \alpha_{\rm N}\bar{f} \right], \nonumber \\
\frac{1}{Nf}\mathcal{L}\left({\bf c},1\right) & = & \alpha_{\rm H} \left((1-\rho_{\rm A}-\rho_{\rm N})\Theta(d({\bf c})) + \rho_{\rm A}q(d({\bf c}),A_{\rm A}) +  \frac{\rho_{\rm N}}{fP}\left(q(d({\bf c}),A_{\rm N})+(fP-1)q(d({\bf c}),\beta_{\rm N}A_{\rm N})\right) -f\right)  \nonumber \\
        & &    + \alpha_{\rm A}\rho_{\rm A}\left(q(d({\bf c}),A_{\rm A}) -\bar{f}\right) + \frac{\alpha_{\rm N}\rho_{\rm N}}{fP}\left(q(d({\bf c}),A_{\rm N})+(fP-1)q(d({\bf c}),\beta_{\rm N}A_{\rm N})-\bar{f}\right).
\end{eqnarray}
Note that now $d({\bf c}) =(1-a) c^E-(\theta +b)$ and that the number of parameters in the reduced model is decreased by 1 ($\alpha_{\rm I}$ still appears in the full model, and we must ensure it is large enough to maintain tight E/I balance).

The interference $\mathcal{N}$ was previously computed in a six-dimensional space [see Supp. Mat. \ref{sec:N}, Eq. (\ref{eq:6D})]. Under the additional constraint of E/I balance the space becomes four-dimensional because of the two linear constraints $r^{EI}_\mu = r^{EI}_\nu = 0$. Eqs. (\ref{eq:N1}, \ref{eq:N20}, \ref{eq:N21}) become now-- 
\begin{eqnarray}
\frac{1}{Nf^2}\mathcal{N}_1({\bf c}) & = & \left[\alpha_{\rm A} \rho_{\rm A}\left(1-\beta_{\rm A}\right) + \alpha_{\rm N} \rho_{\rm N}\right] (\psi(\mathbf{c})-\bar{f}) + \alpha_{\rm H} (\psi(\mathbf{c})-f),\nonumber\\
\frac{1}{Nf^2} \mathcal{N}_{2,0}(\mathbf{c}) & = & \rho_{\rm N}( \alpha_{\rm N} +\alpha_{\rm H}) \tilde{p}(d(c_{\nu}),A_{\rm N})-(\alpha_{\rm N} \rho_{\rm N}\bar{f} + \alpha_{\rm H} f),\nonumber\\
\frac{1}{Nf^2} \mathcal{N}_{2,1}(\mathbf{c})&  =& \frac{\rho_{\rm N}}{fP}(\alpha_{\rm A} \rho_{\rm A}(1-\beta_{\rm A}) + \alpha_{\rm N} + \alpha_{\rm H})\tilde{p}(d(c_{\nu}),A_{\rm N})\nonumber \\
&& - \rho_N (\alpha_{\rm A} \rho_{\rm A} (1-\beta_{\rm A}) + \alpha_{\rm N} ) \bar{f} -\alpha_{\rm H} f ,
\end{eqnarray}
where $\psi({\bf c})$ is defined in Eq. (\ref{eq:psi}) and $d(c_\nu)=(1-a) c^E_\nu-(\theta +b)$.

\begin{table}
	\begin{center}
		\begin{tabular}{|c|l|c|c|c|}
		 \hline
 		 Parameter & Meaning & Value & Value & Value \\
   		 \hline\hline
   		 $b$ & E/I balance line offset & 0.3 & 2.25 & 4.2 \\
   		 \hline
 		 $\alpha_{\rm A}$ & ACh learning rate & 0.04 & 0.8  & 0.8 \\
 	 	 \hline
		 $\rho_{\rm A}$ & ACh pairing rate & 0.05 & 0.15 & 0.15\\
		 \hline
		 $A_{\rm A}$ & ACh disinhibitory amplitude & 1 & 1 & 1 \\
		 \hline
		 $\beta_{\rm A}$ & ACh-dependent LTP/LTD ratio & 3 & 1.6 & 1.4 \\
		 \hline
		 $\alpha_{\rm N}$ & NE learning rate  & 0.12& 0.007 & 0.007\\
		 \hline
		 $\rho_{\rm N}$ & NE pairing rate & 0.001& 0.001& 0.001\\
		 \hline
		 $A_{\rm N}$ & NE disinhibitory amplitude & 5  & 5& 5 \\
		 \hline
		 $\alpha_{\rm H}$ & Hebbian learning rate & $\{0,10^{-5},5\cdot 10^{-5},10^{-4},5\cdot 10^{-4}\}$& 0& 0\\
		 \hline
		 $\alpha_{\rm I}$ & Inhibitory learning rate & 0.3 & 0.45& 0.45\\
		 \hline\
		 $r_{\rm A}$ & Update rate of $\rho_{\rm A}$ (Eq. \ref{eq:DrhoA}) & 0.002 & 0.006& 0.006\\
		 \hline\
		 $r_{\rm N}$ & Update rate of $\rho_{\rm N}$ (Eq. \ref{eq:DrhoN}) & 0.004 & 0.012& 0.012\\
		 \hline  
		 $[\rho^{\min}_{\rm A},\rho^{\max}_{\rm A}]$ & Allowed range for $\rho_{\rm A}$ & $[0.05,\,5]\,\rho_{\rm A}$ & $[0.005,\,5]\,\rho_{\rm A}$& $[0.005,\,5]\,\rho_{\rm A}$\\
		 \hline  
		 $[\rho^{\min}_{\rm N},\rho^{\max}_{\rm N}]$ & Allowed range for $\rho_{\rm N}$ & $[0.05,\,5]\,\rho_{\rm N}$ & $[0.005,\,5]\,\rho_{\rm N}$& $[0.005,\,5]\,\rho_{\rm N}$\\
		 \hline  
		\end{tabular} 
	\end{center}
	\caption{Model parameters in the case of E/I balance line tilted ($a=0.7$) with respect to the spiking threshold. Fixed model parameter: $\bar{f} = 0.01$ (reference spiking level for neuromodulatory plasticity). Task parameters used $f=0.2$, $N_E = 3200,\,N_I = 800$, $\theta = 1$.}		\label{tab:redmodoptimtilt}
\end{table}

\section{Approximations in the reduced model}\label{sec:redapprox}

The derivation of the reduced model and the optimisation of model parameters (Supp. Mat. \ref{sec:L}, \ref{sec:N}, \ref{sec:numoptim}, \ref{sec:LNb}) are  based on a number of approximations and simplifying assumptions that make the connection between the reduced and full models imperfect. 

{\bf Uniform distribution of associations in reduced weight space.} Optimisation of model parameters is performed by maximising the ratio of ``good'' vs. ``bad'' weight updates, i.e., $\mathcal{L}(\mathbf{c},Y)/\mathcal{N}(\mathbf{c})$ averaged over the reduced weight space (Supp. Mat. \ref{sec:numoptim}). To carry out this average numerically, we must make an assumption on the distribution of associations in the reduced weight space. We assume this distribution to be uniform, but as the learning process unfolds, associations become concentrated in specific areas of the weight space. In the example shown in Fig. \ref{fig:full}a, two inputs corresponding to a target spike are misclassified (i.e., these inputs do not lead to a spike, see open red circles). The reduced model for this parameter predicts that these misclassified inputs should be driven across  the spiking threshold (black), even in the presence of the remaining associations. However, the remaining associations are clearly not uniformly distributed in the reduced weight space, so the reduced model does not accurately capture the learning dynamics in this case. This discrepancy between the reduced and full model is partially alleviated when dynamics are restricted to the E/I balance line (Figs. \ref{fig:reduced}b, \ref{fig:full}c, Supp. Mat. \ref{sec:LNb}), because the dimension of the reduced E-I space is smaller (one instead of two for $\mathcal{L}$, and four instead of six for $\mathcal{N}$). 

{\bf Sign-constrained synapses.} The derivation of the reduced model does not take into account the sign-constraint of individual synaptic weights which are summed to give the E/I currents $c^E, c^I$. In practice, the plasticity mechanisms depress a large number of weights to 0, and in the absence of the constraint would lead to negative weights.

{\bf Finite size effects of input overlaps.} The derivation of the interference function $\mathcal{N}({\bf c})$ assumes that the overlap of inputs $X_\mu \cdot X_\nu$ is equal to its mean in the limit $N\to\infty$. In practice the overlaps vary, with standard deviation that goes to 0 slowly ($\sim 1/\sqrt{N}$), so we expect the finite input dimension to degrade the correspondence between the reduced and full model.

{\bf Parameter scaling with problem size.} The stimulus specificity of the neuromodulatory pairing and disinhibition statistics differs between the two mechanisms we consider (ACh and NE). Therefore we expect that, unless some model parameters depend on the size of the learning task ($P$), the balance between the effects of ACh and NE will break as $P$ changes. Furthermore, since the reduced model becomes intractable for $P>2$, it cannot be used to extract relationships between parameters controlling the neuromodulatory mechanisms that would lead to successful learning for arbitrary problem sizes. In the main text we therefore propose two strategies to find parameters leading to successful learning. 

\section{Dynamic updates of neuromodulation pairing rates} \label{sec:rhot}
We showed in the main text that, for moderate input dimensionality ($N=N_E+N_I=1000$), the values of the neuromodulation pairing rates that lead to successful learning of a task depend on the overlap of excitatory inputs that should lead to an output spike (i.e., ``good'' $\rho_{\rm A},\,\rho_{\rm N}$ depend on $r^E_{11} = \langle X^E_{\mu} \cdot X^E_{\nu} | Y_\mu = Y_\nu = 1 \rangle$). Instead of searching over a range of $\rho_{\rm A},\,\rho_{\rm N}$ values numerically, we showed that update rules for the pairing rates could be used to find values that work for a given task. 

The update rule for ACh is:
\begin{eqnarray}
\Delta \rho_{\rm A} & = &  r_{\rm A} s\left(\hat{\varepsilon}_0-\hat{\varepsilon}_1\right),  \label{eq:DrhoA}\\
\hat{\varepsilon}_0 & = & \frac{1}{fP}\sum_{\mu=1}^P \left(1-Y_\mu\right)\left|y_\mu-Y_\mu\right|,\nonumber\\
\hat{\varepsilon}_1 & = & \frac{1}{(1-f)P}\sum_{\mu=1}^P Y_\mu\left|y_\mu-Y_\mu\right|.\nonumber
\end{eqnarray}
Here, $\hat{\varepsilon}_1,\,\hat{\varepsilon}_0$ are the normalised errors corresponding to inputs with spike/no-spike target outputs, respectively. The update rule in Eq. (\ref{eq:DrhoA}) searches for $\rho_{\rm A}$ that balances the rate of erroneous responses when the target is a spike/no-spike. Since ACh-dependent plasticity is biased towards depression in our model, $\hat{\varepsilon}_0>\hat{\varepsilon}_1$ (i.e., more erroneous spikes than erroneous no-spikes) corresponds to an {\em increase} in ACh pairing rate which then leads to stronger LTD of excitatory synapses. 

In some cases, especially when the number of associations is small, a range of $\rho_{\rm A},\,\rho_{\rm N}$ are suitable for a single classification problem. To promote solutions with a low pairing rate, the function $s(\cdot)$ [Eq. (\ref{eq:DrhoA})] preserves the sign of its argument, but the slope on the negative side (decreasing the pairing rate) is larger than on the positive side: 
\begin{equation}
s (x) = \begin{cases}
x & x\ge 0\\
3x & x<0 
\end{cases},
\end{equation}
From inspection of a large number of numerical simulations, we find that if the slope on the negative side is much larger than 3, the learning process can get stuck with ACh pairing rate close to 0. When the slope is close to 1, solutions with low pairing rates are not strongly favoured. 

The update rule for NE is:
\begin{eqnarray}
\Delta \rho_{\rm N} & = &   \frac{r_{\rm N}}{P}\sum_{\mu=1}^P \left(f-y_\mu\right).  \label{eq:DrhoN}
\end{eqnarray}
The update rule in Eq. (\ref{eq:DrhoN}) searches for $\rho_{\rm N}$ where the average response is equal to the average target response. Since NE pairing leads to overall LTP of excitatory weights, if (on average) $f>y_\mu$, then $\rho_{\rm N}$ should increase.  
Updates of $\rho_{\rm A},\,\rho_{\rm N}$ occur after every cycle over $P$ inputs is completed, but are much slower than updates of synaptic weights. Evidence that the mid-brain ACh and NE update their activity levels based on the animal's performance, consistent with Eqs. (\ref{eq:DrhoA}, \ref{eq:DrhoN}) is cited in the main text. Finally, we note that in simulations $\rho_{\rm A}$ is bounded from below and from above by $\rho^{\min}_{\rm A},\,\rho^{\max}_{\rm A}$ (and similarly for $\rho_{\rm N}$ but no saturation to the upper bound is seen in cases where learning is successful (not shown).  

% \begin{equation}
% \Delta \rho_{\rm A} = \frac{r_{\rm A}}{P} \sum_{\mu=1}^P \left[ \frac{1-Y_\mu}{}\right]
% \end{equation}

\section{Capacity under detailed E/I balance}\label{sec:BalanceCapacity}

Our analysis of the proposed learning algorithm shows that robust learning of a large number of input--output associations requires {\em detailed} E/I balance. \cite{Rubin2017} studied the encoding capacity of a perceptron with excitatory and inhibitory weights [Eq. (\ref{eq:EIperceptron}), excluding the neuromodulatory terms $M_{\rm A}$, $M_{\rm N}$). These authors showed that, to overcome input noise (e.g., failure of  $X_{\mu,i}=1$ inputs to spike) {\em and} output noise (e.g., a spurious output spike despite the net current being below threshold), solutions must maintain {\em global} E/I balance. This form of balance is formally defined through the sum of all E, I synaptic weights being of order $\sqrt{N_E},\,\sqrt{N_I}$. Interestingly, the reduction in capacity due to the global balance requirement is typically small, unless excitatory inputs substantially outnumber inhibitory inputs.

Requiring synaptic weights to maintain {\em detailed} balance further reduces the capacity because for each input--output association $\mu$, the weights now must satisfy,
\begin{equation}
\left(\sum_{i=1}^{N_I} w^I_i X_{\mu,i}^I\right) = a \left(\sum_{i=1}^{N_E} w^E_i X_{\mu,i}^E\right) + b\label{eq:tiltedbalance}
\end{equation}
(in addition to correct encoding of the association itself). We are unsure whether the replica theory that \cite{Rubin2017} used to compute the capacity of an E, I perceptron with global balance can be extended to apply to the scenario with detailed balance. The reason is that the properties of synaptic weight satisfying the input--output associations are studied through {\em order parameters} where excitation and inhibition is on equal footing, as it is in Eq.~(\ref{eq:EIperceptron}). In our case, we would still need the order parameters used in \cite{Rubin2017}, and additionally need order parameters where sums of E, I weights are performed {\em not} on equal footing [see Eq. (\ref{eq:tiltedbalance}) where $a<1$]. 

Further investigation into a possible exact analytical solution for the capacity of a perceptron with detailed balance is beyond the scope of this paper. We are however able to obtain an upper bound on the capacity. We do so by substituting the sum over inhibitory weights in Eq. (\ref{eq:EIperceptron}) with the right-hand-side of Eq. (\ref{eq:tiltedbalance}) which gives (excluding the neuromodulation terms) 
\begin{equation}
y = \Theta \left( \sum_{i=1}^{N_E} w^E_i X_{\mu,i}^E - \frac{\theta+b}{1-a} \right).\label{eq:Eperceptron}
\end{equation}
This is the definition of a {\em sign constrained perceptron} which was analysed in \citep{Amit1989b, Nadal1990, Brunel2004} and for which the capacity is known. These results give only an upper bound for the capacity of an E, I perceptron with detailed balance because in substituting the sum over inhibitory weights to get Eq. (\ref{eq:Eperceptron}) we ignored the sign constraint on these weights. In other words, Eq. (\ref{eq:tiltedbalance}) must be satisfied with $w^I_i>0$, a restriction which is not taken into account in Eq. (\ref{eq:Eperceptron}) and thus the true capacity is bounded from above by the existing results for a sign-constrained perceptron.

\section{Fitting inhibitory plasticity parameters to data} \label{sec:Ifit}

We begin by assuming the following mathematical form for $\Delta I$, 
\begin{equation}
\Delta I = \frac{G C_E + K}{C_I}- \alpha_{\rm I}.\label{eq:DwIexp1}
\end{equation}
Setting $K=0$ reduces to the fit done in \cite{Damour2015} for a linear dependence of inhibitory plasticity on the E/I ratio before pairing. Note that there the data was not pooled over pre$\to$post and post$\to$pre protocols. Defining 
\begin{equation}
A = G/\alpha_{\rm I},\qquad B=K/\alpha_{\rm I},
\end{equation} 
Eq.~(\ref{eq:DwIexp1}) can be rewritten as Eq. (\ref{eq:DwIexp2}) in the main text, which we recognised as a nonlinear version of Eq. (\ref{eq:DwI}).  Our theory makes the following predictions pertaining to the fit parameters.

We fit the parameters $C,K,\alpha_{\rm I}$ (Eq. \ref{eq:DwIexp1}) to the data triplets $\{c^E,\,c^I,\,\Delta I\}$ reported in \cite{Damour2015} (see brief description of the part of the experiment relevant to the current work in the main text). We performed a least squares fit twice: once over the entire dataset ($n=49$ neurons), and excluding 7/49 outliers with large ($>300$pA) EPSCs or IPSCs. The full (excluded) dataset contains 8/49 (6/42) points for which the pairing led to iLTD ($\Delta I<0$). In each fit, we chose different weights for iLTP/iLTD points, such that the number of points in each group times the weight is equal for iLTP, iLTD. We performed two additional fits with $K$ fixed to 0 (see Eq. \ref{eq:DwIexp1}). This recovers the form of the fit done in \cite{Damour2015}, where $\Delta I$ depends on the EPSC and the IPSC only through their ratio. The results of the fits are summarised in Table \ref{tab:Ifit}.

\begin{table}
\noindent \centering{}%
\begin{tabular}{|c|c|c|c|c|}
\hline 
Fit & $K=0$, excl. outliers & $K$ free, excl. outliers & $K=0$, incl. outliers & $K$ free, incl. outliers\\
\hline 
\hline 
$\alpha_{\rm I}$ & 0.124 (0.069, 0.17) & 0.046 (-0.033, 0.12) & 0.126 (0.077, 0.17) & 0.0759 (0.0028, 0.14)\\
\hline 
$G$ & -0.1296 (-0.34, 0.089) & -0.2057 (-0.41, 0.0086) & -0.08 (-0.27, 0.10) & -0.126 (-0.31, 0.062)\\
\hline 
$K$ & N/A & 9.9 (2.0, 17) & N/A & 6.69 (-0.56, 13)\\
\hline 
$R^{2}$ & 0.340 & 0.435 & 0.366 & 0.410\\
\hline 
$RMSE$ & 0.251 & 0.236 & 0.259 & 0.252\\
\hline 
\end{tabular}\caption{\label{tab:Ifit}Inhibitory plasticity fitting results.}
\end{table}

We note that in \citep{Damour2015}, the holding potential used to measure IPSCs was lower than in other studies ($-40$ compared to $0-10$ mV). Using a higher potential in the voltage-clamp mode would have lead to an overall increase in the measured IPSCs. The resulting plasticity however would not have changed, since induction was done in current-clamp mode. Despite the dependence of the currents on the holding potential, we expect that our main result here--the slope of the E/I balance line--would not be strongly affected, had the holding potential for inhibition been higher. The reason is that {\em all} data points in Fig. \ref{fig:experiment} would have been shifted upwards. This would have surely affected the intersction of the E/I balance line (magenta) and the EPSC=IPSC line (brown), but we expect the effect on the slope of the E/I balance line would have been minor.

\section{Analysis of ACh-dependent plasticity data} \label{sec:LTDbias}

We showed in the main text that the sign and magnitude of plasticity of EPSCs measured in auditory cortex is consistent with our theoretical prediction for ACh-dependent plasticity. Specifically, changes to EPSCs following pairing are a decreasing function of a proxy measure for stimulus overlap ($D^E$, the absolute difference of normalised EPSCs between the stimulus paired with ACh stimulation and every other stimulus). To rule out the possibility that the change in excitation reported previously in \citep{Froemke2013} is a regression to the mean effect wherein small (large) EPSCs undergo LTP (LTD) we did the following analysis. First we computed the correlation between the (normalised) excitatory current before the pairing $\hat{c}^E$ and $D^E$. This correlation is small ($R^2 = 0.111$, Fig. \ref{fig:experiment2}). Second, we linearly regressed $\Delta^E_{\rm A}$ against $\hat{c}^E$ and $D^E$ simultaneously and found that the dependence on $D^E$ predicted by our model is more than three times stronger than the dependence on $\hat{c}^E$, which could be argued to be caused by regression of EPSCs to a mean.

\section{Supplementary discussion of connections with related literature}\label{sec:supdisc}

\paragraph{Perceptron with segregated E, I inputs}
Recent theoretical studies of a perceptron with segregated excitatory (E) and inhibitory (I) inputs yield accurate predictions for cortical synaptic weight distributions \citep{Brunel2016}; and explain how E/I balance leads to noise robustness \citep{Rubin2017}, consistent with empirical signatures of E/I balance \citep{Okun2008, Barral2016}. Importantly, these works do not address the question of how the cortex finds synaptic weights associating inputs with outputs, which is the primary focus of our work.

\paragraph*{Connection to the inhibition-stabilised-network.}

Recent work showed that the Supralinear-Stabilised-Network (SSN) model captures characteristics of cortical activity \citep{Hennequin2018}, including contrast dependence of the E/I ratio in visual cortex \citep{Adesnik2017}. It is notable that the our model makes a similar and complementary prediction (i.e., the tilt of the E/I balance line), despite that \cite{Hennequin2018} focus on network dynamics, while we do not include temporal dynamics in our model and focus instead on learning. 

\paragraph*{Inhibition as prediction.}

How does the cortex compute predictions of high-dimensional stimulus features \citep{Keller2018}? Clearly not every feature is concurrently extracted from the external world and predicted by an internal model. We speculate that mechanisms such as E/I balance promoting a conserved approximate relationship between two input classes can compute a prediction for many stimuli, relying on biophysical constraints rather than a model-based estimates. The direct link our model makes to synaptic plasticity suggests how prediction errors may be used for learning. Extensions of our work will include laminar organisation and inhibitory cell-types, potentially integrating these into the predictive coding framework \citep{Rao1999,Whittington2017}, following evidence that inhibition of layer 4 neurons by superficial layers receiving feedback projections is an important signal for plasticity, learning and memory \citep{Letzkus2015,Froemke2015}.

\paragraph*{Excitatory and inhibitory synaptic weight distributions.} 

At capacity, sign-constrained synapses in a perceptron model follow truncated Gaussian distributions \citep{Brunel2004,Brunel2016,Rubin2017}. Synaptic weights that are equal to 0 were interpreted as silent-- i.e., a synaptic contact that does not lead to a postsynaptic current \citep{Kerchner2008}; or as a potential synapse-- i.e., two neurons that are in close proximity to one another but are nevertheless not connected \citep{Brunel2004,Brunel2016}. Synapses in our model still follow truncated Gaussian weight distributions, to a very good approximation, for both E and I (Fig. \ref{fig:full}f). While previous work focused only on E synapses, our work highlights the important role inhibitory synapses and plasticity to play in learning, posing the question: What is the appropriate interpretation of {\em inhibitory} synaptic weights equal to 0, and does the theoretical prediction match experiments? Some data regarding the density and organisation of inhibitory connections does exist \citep{Katzel2011}; as well as evidence for silent inhibitory synapses in rat hippocampus and goldfish \citep{Tong1996,Charpier1995}. We expect that in the future, high-throughput experimental techniques for measuring synaptic weight distributions will help answering this question. Such data will also further constrain theories of learning and plasticity \citep{Barbour2007}, especially in models that include inhibitory plasticity.

\newpage

\setcounter{figure}{0}    
\renewcommand{\thefigure}{S\arabic{figure}}

\begin{figure}[h!]
	\begin{center}
		\includegraphics[scale=0.43]{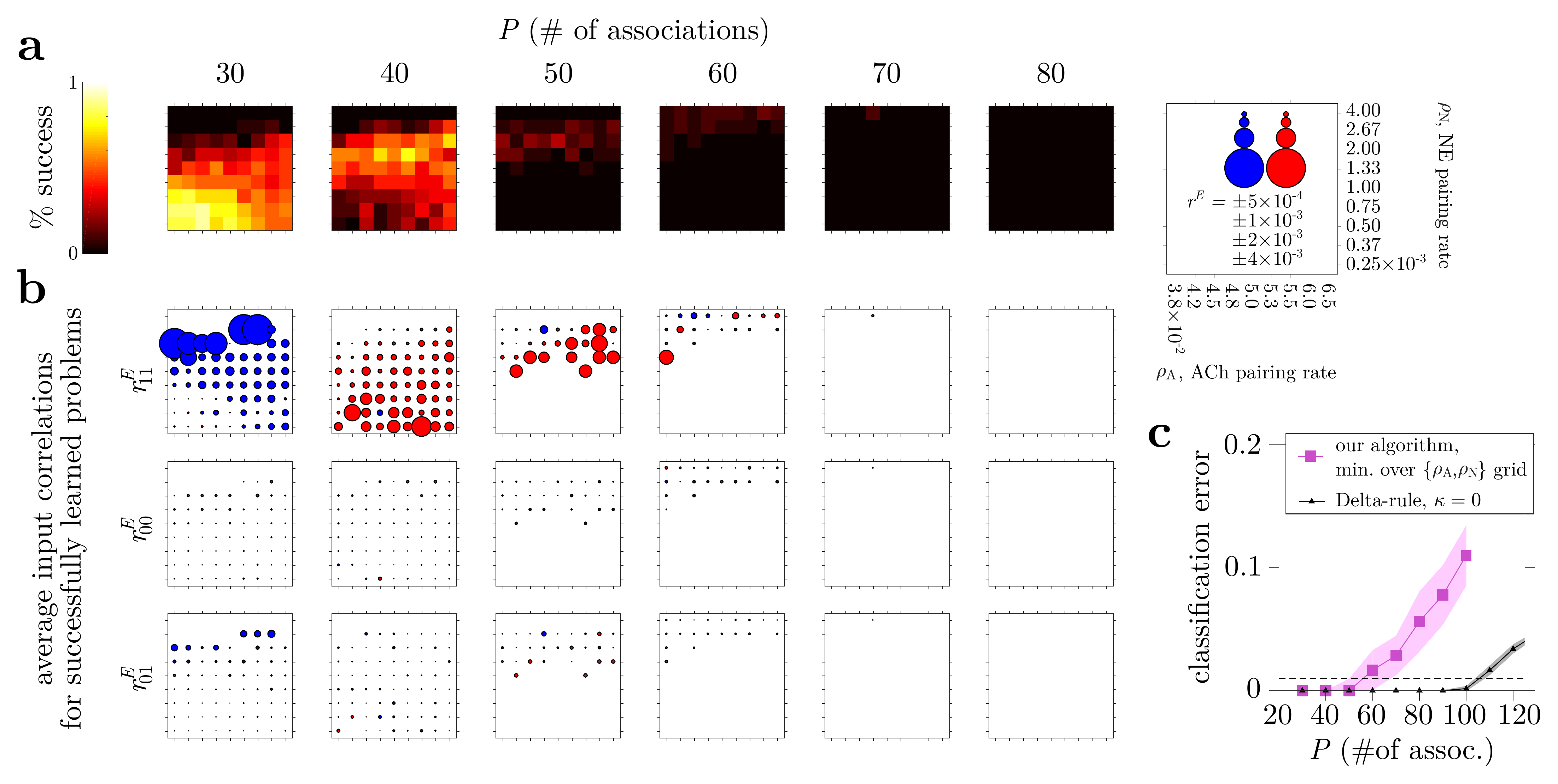}
	\end{center}
	\caption{{\bf Finite size effects of input correlations.} We plot results for a model with the same parameters as Fig. \ref{fig:full}, Table \ref{tab:redmodoptimtilt} (left column), with two modifications: smaller input dimensionality ($N_E=800,\,N_I=200$, compared to $N_E=3200,\,N_I=800$ throughout the paper), and fixed $\rho_{\rm A},\,\rho_{\rm N}$. Performance is shown across a 9$\times$9 grid of values of the ACh and NE pairing rates $(\rho_{\rm A},\rho_{\rm N})$. The legend on the right shows the $\rho$ values used. ({\bf a} We plot the percent success, defined to be the fraction of simulations with final classification error $<1\%$ as a function of the two $\rho$ values, for multiple problem sizes. The values of $\rho$ that lead to good performance depend on $P$ systematically, and in correlated fashion. ({\bf b}) For each problem that was learned successfully by a model with a specific combination of $\rho$ values, we computed the average excitatory input correlations separately for pairs of inputs corresponding to target $\{0,0\}$, $\{0,1\}$, and $\{1,1\}$ outputs (denoted $r^E_{00},\,r^E_{01},\,r^E_{11}$, respectively). Red/blue circles correspond to positive/negative correlation, and the size corresponds to the magnitude of the average correlation (legend shown in top right). The top row (correlations between pairs of inputs that both correspond to a target output spike) shows that different combinations of $\rho$ are needed, depending on $r^E_{11}$ of a given problem. Input correltions go to their average values slowly as $N$ increases. Therefore, to learn any problem of a finite size (with correlations that differ from the mean), the $\rho$'s must be dynamically adjusted. ({\bf c}) The performance of our algorithm taking the minimum over all combinations of $\rho$ for each task simulated, compared with the Delta-rule. Searching over a grid of $\rho$ values leads to similar performance to the algorithm with dynamic updates of $\rho$ and to the Delta-rule, once the margin for noise $\kappa$ is taken into account (see Fig. \ref{fig:full}, Fig. \ref{fig:Kappa}).}\label{fig:rho}
\end{figure}

\newpage

\begin{figure}[h!]
	\begin{center}
		\includegraphics[scale=0.43]{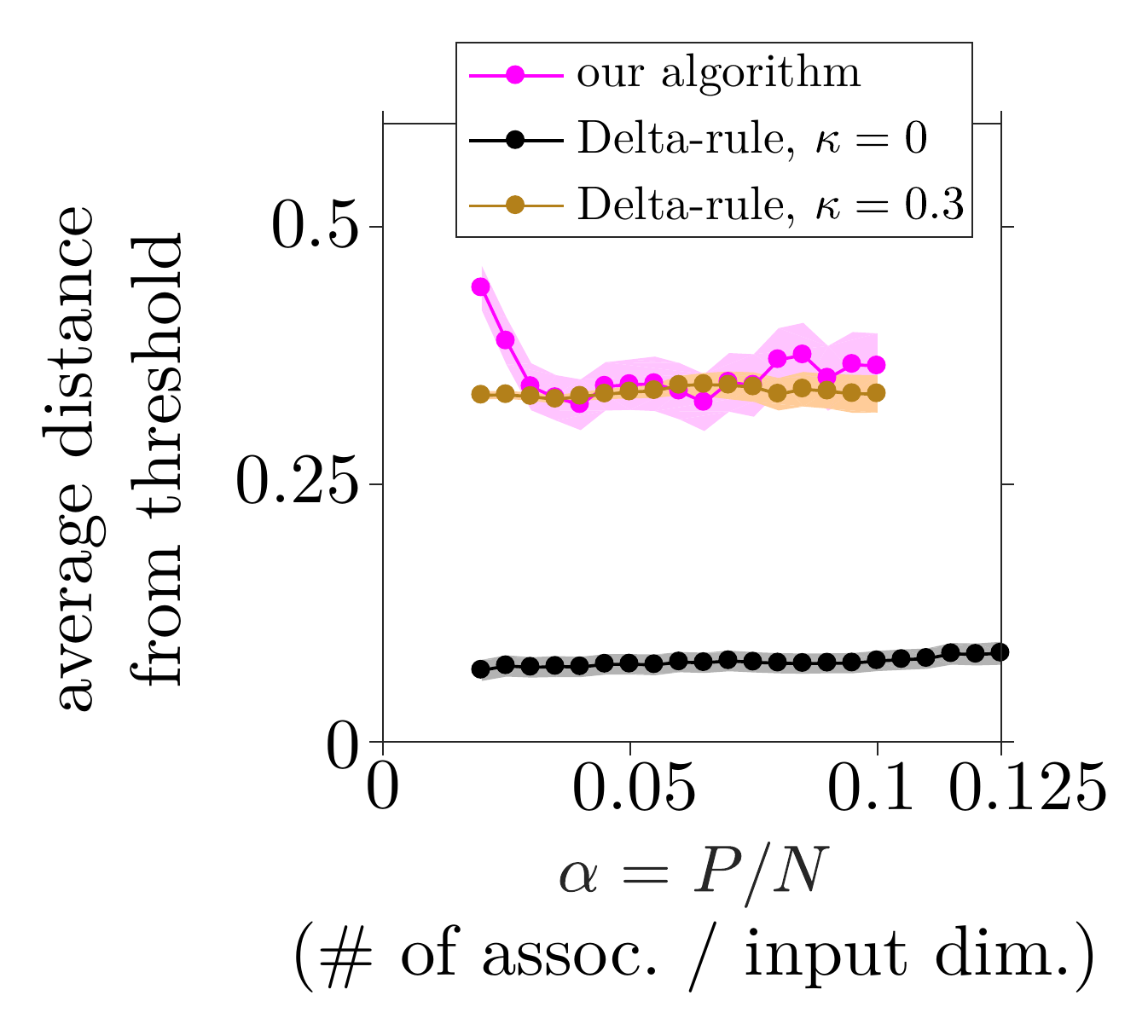}
	\end{center}
	\caption{{\bf Matching the Delta-rule margin $\bm{\kappa}$ to our algorithm.} We plot the average distance of correctly classified inputs from the spiking threshold, as a function of the problem size $\alpha = P/N$. Our algorithm, even when the Hebbian learning rate is 0 ($\alpha_{\rm H}= 0$, magenta) results in a margin for noise more than 3 times larger than the Delta-rule when the margin parameter is 0 ($\kappa=0$, black). When the margin parameter is set to $\kappa=0.3$, the resulting average margin matches our algorithm over a range of $\alpha$ values.}\label{fig:Kappa}
\end{figure}

\newpage

\begin{figure}[h!]
	\begin{center}
		\includegraphics[scale=0.43]{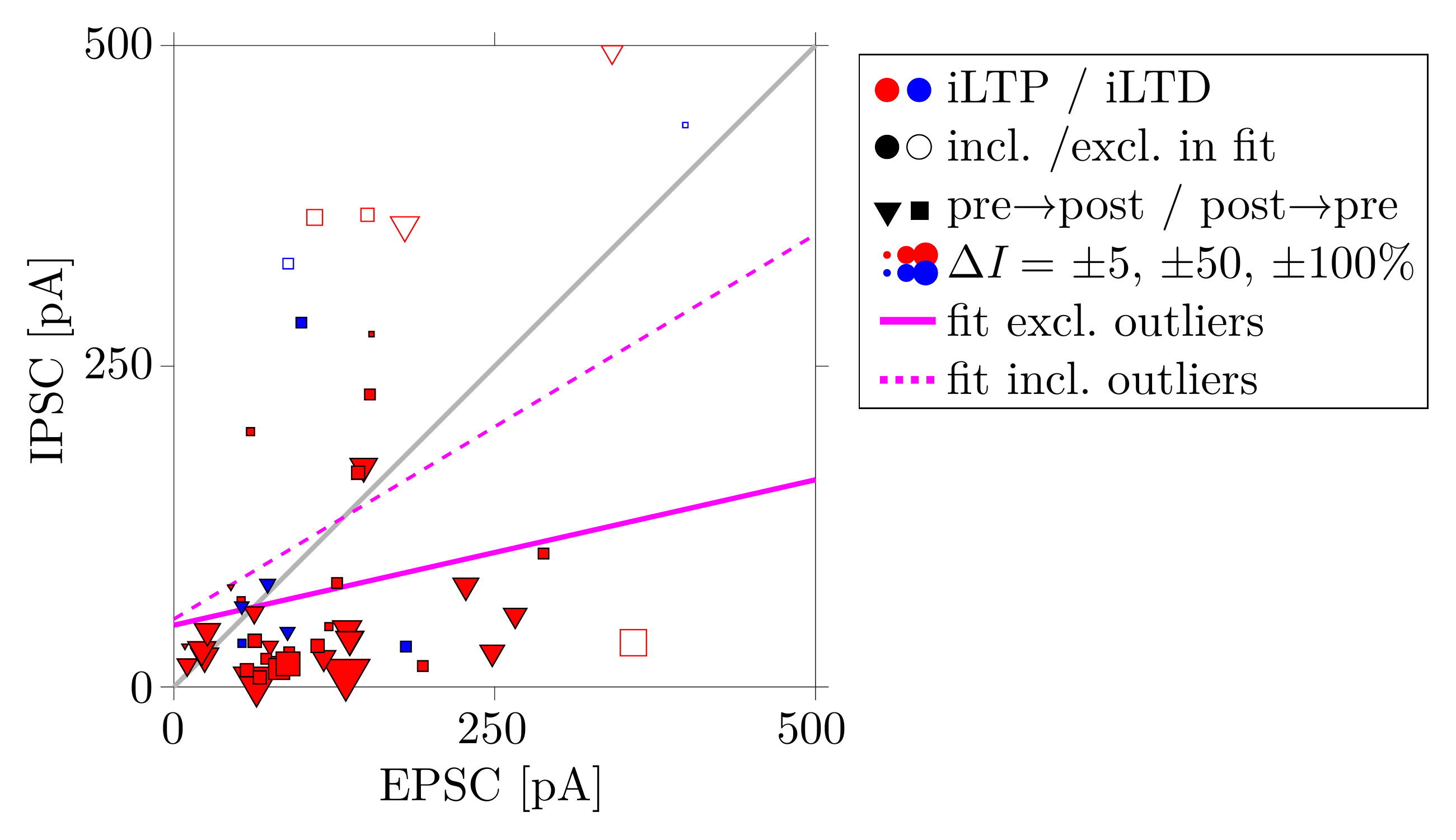}
	\end{center}
	\caption{{\bf Validation of theoretical predictions: inhibitory plasticity measured {\em in vitro} \citep{Damour2015} drives towards a tilted E/I balance line.} Same as Fig. \ref{fig:experiment}, showing fits of the E/I balance line excluding points with EPSC or IPSC $>300$pA (solid magenta line), and including all points (dashed magenta line). Excluded points are indicated by open symbols. The slope of the E/I balance line when fitting using all points is $<1$, as predicted by our theory. However a slope of 1 cannot be ruled out with $>95\%$ confidence. The slope ($A=0.64$) is similar to the one used throughout the paper in models with a tilted balance line ($a=0.7$).}\label{fig:tilt2}
\end{figure}

\newpage

\begin{figure}[h!]
	\begin{center}
		\includegraphics[scale=0.4]{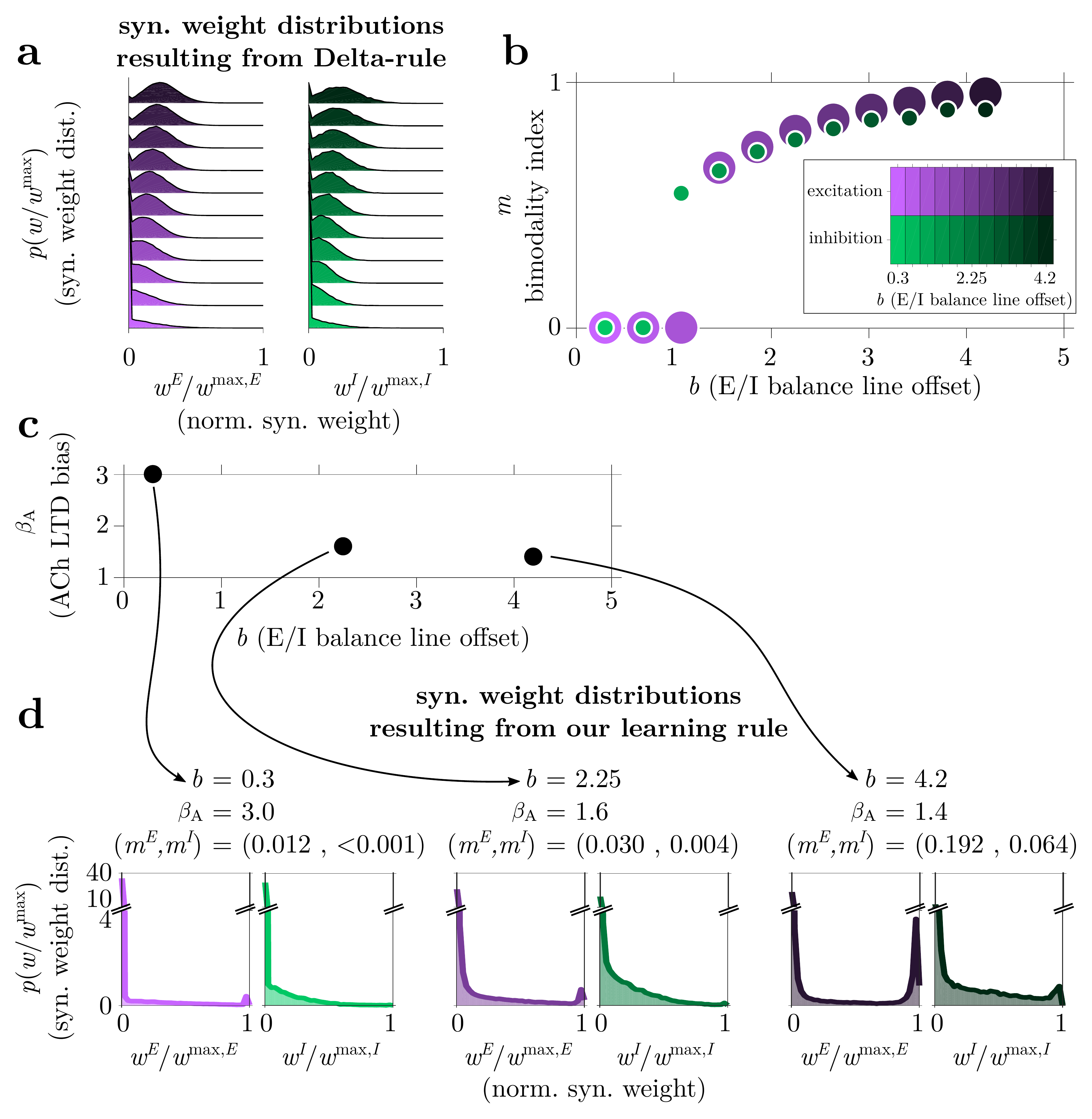}
	\end{center}
	\caption{{\bf Effect of E/I balance line offset on synaptic weight distributions and on ACh LTD bias.} ({\bf a}) Excitatory (purple) and inhibitory (green) synaptic weight distributions resulting from learning classification problems with $P = 140$, $N=4000$ using the Delta-rule with detailed balance (margin $\kappa=0.3$). The offset of the E/I balance line ($b$) was varied between 0.3-4.2 (light to dark). For all values of $b$, a fraction of synapses is saturated to 0, the lower bound. As $b$ increases, this fraction decreases, and distributions becomes more bimodal. ({\bf b}) We plot the bimodality index of the weight distributions in (a), defined to be the integral over the distribution from the anti-mode (i.e., the minimum between the two peaks) to the upper bound. ({\bf c}) For three example values of the offset, we find parameters for our learning rule resulting in successful classification of $P=140$ inputs. For values of $b$  yielding a unimodal distribution (using the Delta-rule, panels a, b), the parameter $\beta_{\rm A}$ (the ratio of LTD of non-paired inputs relative to LTP of paired inputs) must be significantly larger than one (here, $\beta_{\rm A} = 3$). ({\bf d}) We plot the synaptic weight distributions resulting from our learning rule, with the parameter $\beta_{\rm A}$ shown in (c). Similarly to the Delta-rule, the weight distributions become more bimodal for larger offset. Above each panel we indicate the values of the offset $b$, the ACh LTD bias parameter $\beta_{\rm A}$, and the bimodality indices $(m^E,m^I)$ (computed separately for the E and I weight distributions). Only models with small offset result in unimodal distributions, as seen in experiments. These require an LTD bias of ACh plasticity when learning using our algorithm, thus we predict that ACh-dependent plasticity will be biased towards depression. }\label{fig:offset}
\end{figure}

\newpage

\begin{figure}[h!]
	\begin{center}
		\includegraphics[scale=0.43]{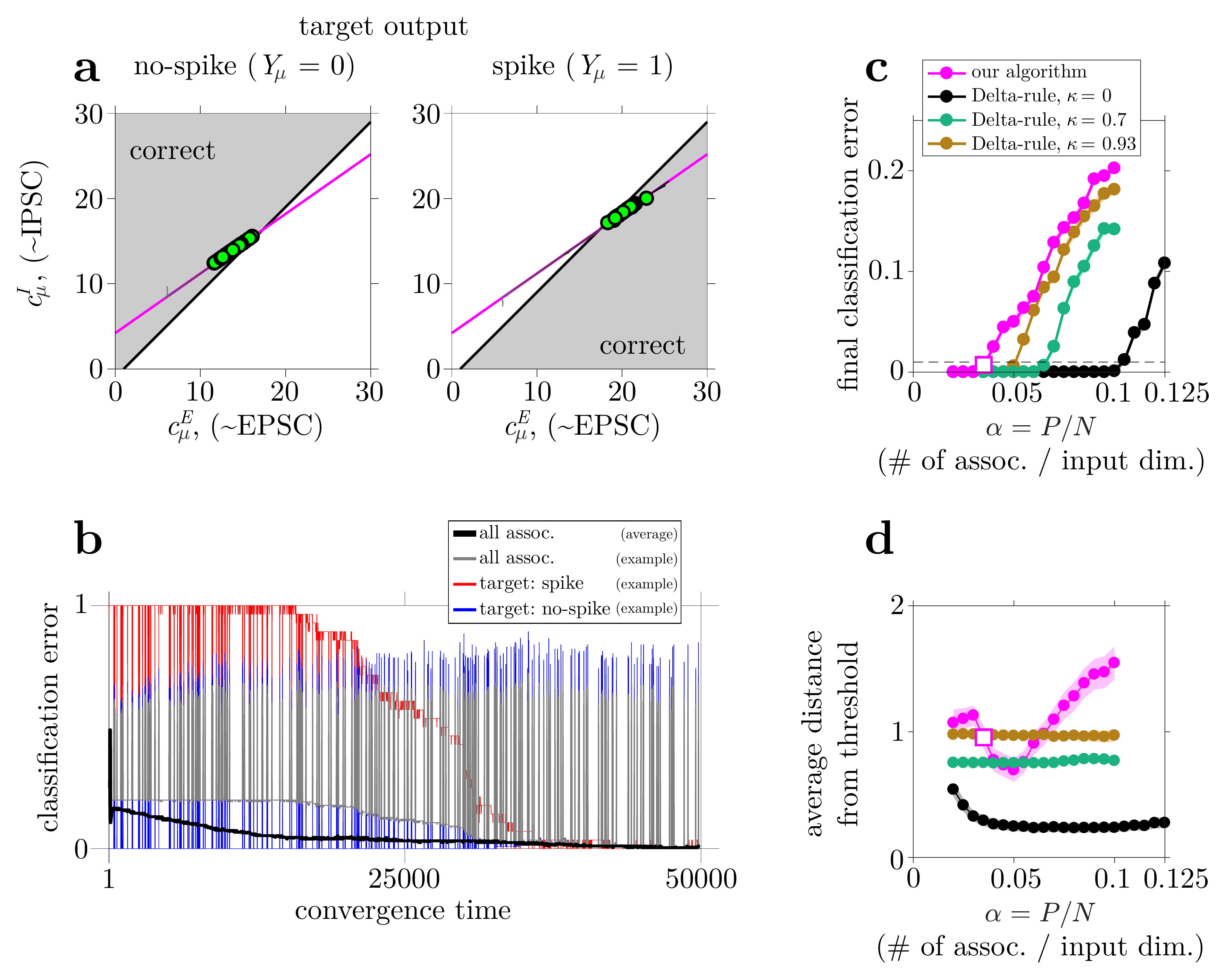}
	\end{center}
	\caption{{\bf Robust learning when the E/I balance line and the spiking threshold intersect far from the origin (EPSC=ISPC=0).} ({\bf a,b}, same as (c,d) in Fig. \ref{fig:full}) Example learning dynamics of full model projected down to the reduced E-I space. When the offset of the E/I balance line is large ($b=4.2$) the E/I balance line intersects the spiking threshold far from the origin [compare position of intersection of E/I balance line with spiking threshold here ($v^E,v^I\approx 15$) with its position in Fig. \ref{fig:full} ($v^E,v^I\approx 4$)]. In this scenario our algorithm with modified parameters (see Table \ref{tab:redmodoptimtilt}) is still able to learn the classification problem. ({\bf c}, same as (e) in Fig. \ref{fig:full}) We plot the classification error as a function of the problem size $\alpha = P/N$. Here the average margin of correctly classified inputs varied with $P$ (shown in panel d), so we compared to the learning performance using the Delta-rule using two different values of the margin $\kappa$ ($\kappa = 0.7$, green; $\kappa=0.93$, brown). Our algorithm can learn problems up to $\sim 70\%$ of the theoretical capacity. This definition of the model results in bimodal synaptic weight distributions (see Fig. \ref{fig:offset}).}\label{fig:offset2}
\end{figure}

\newpage

\begin{figure}[h!]
	\begin{center}
		\includegraphics[scale=0.43]{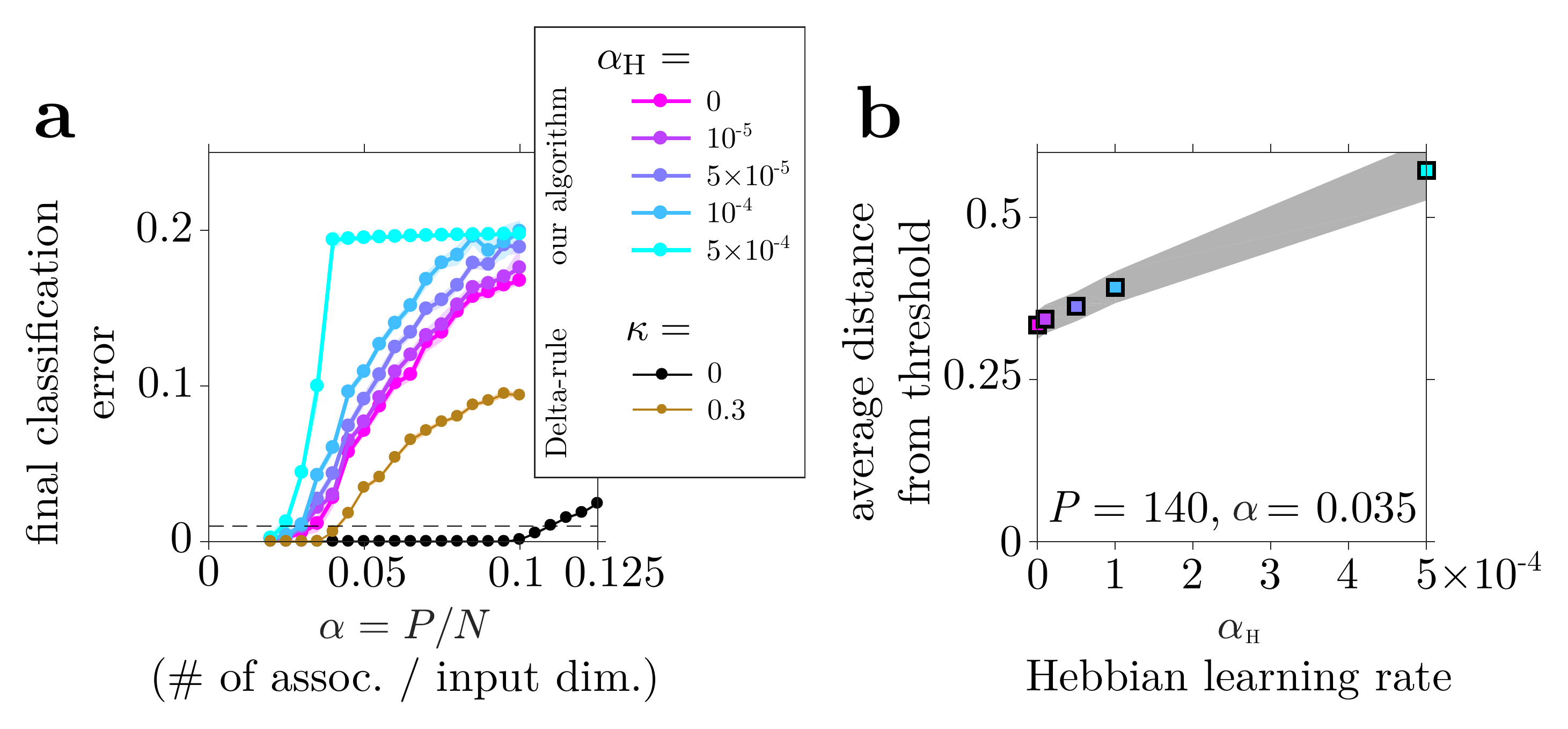}
	\end{center}
	\caption{{\bf Effect of Hebbian learning rate on performance and noise robustness.} ({\bf a}) We plot the classification error as a function of the problem size $\alpha = P/N$ for our algorithm with (parameters shown in the left column of Table \ref{tab:redmodoptimtilt}) for a number of values of the Hebbian learning rate $\alpha_{\rm H}$. Also shown for comparison the results of learning using the Delta-rule using two values of the margin parameter $\kappa$. As $\alpha_{\rm H}$ increases, the algorithm's ability to robustly learn to classify a large number of inputs deteriorates, because  $\alpha_{\rm H}$ controls target-response-independent learning which interferes with the neuromodulatory mechanisms. ({\bf b}) The average distance of correctly classified inputs from the spiking threshold, as a function of $\alpha$. Increasing $\alpha_{\rm H}$ results in larger robustness to noise, since Hebbian plasticity drives inputs away from the spiking threshold.}\label{fig:alphaHebb}
\end{figure}

\newpage

\begin{figure}[h!]
	\begin{center}
		\includegraphics[scale=0.43]{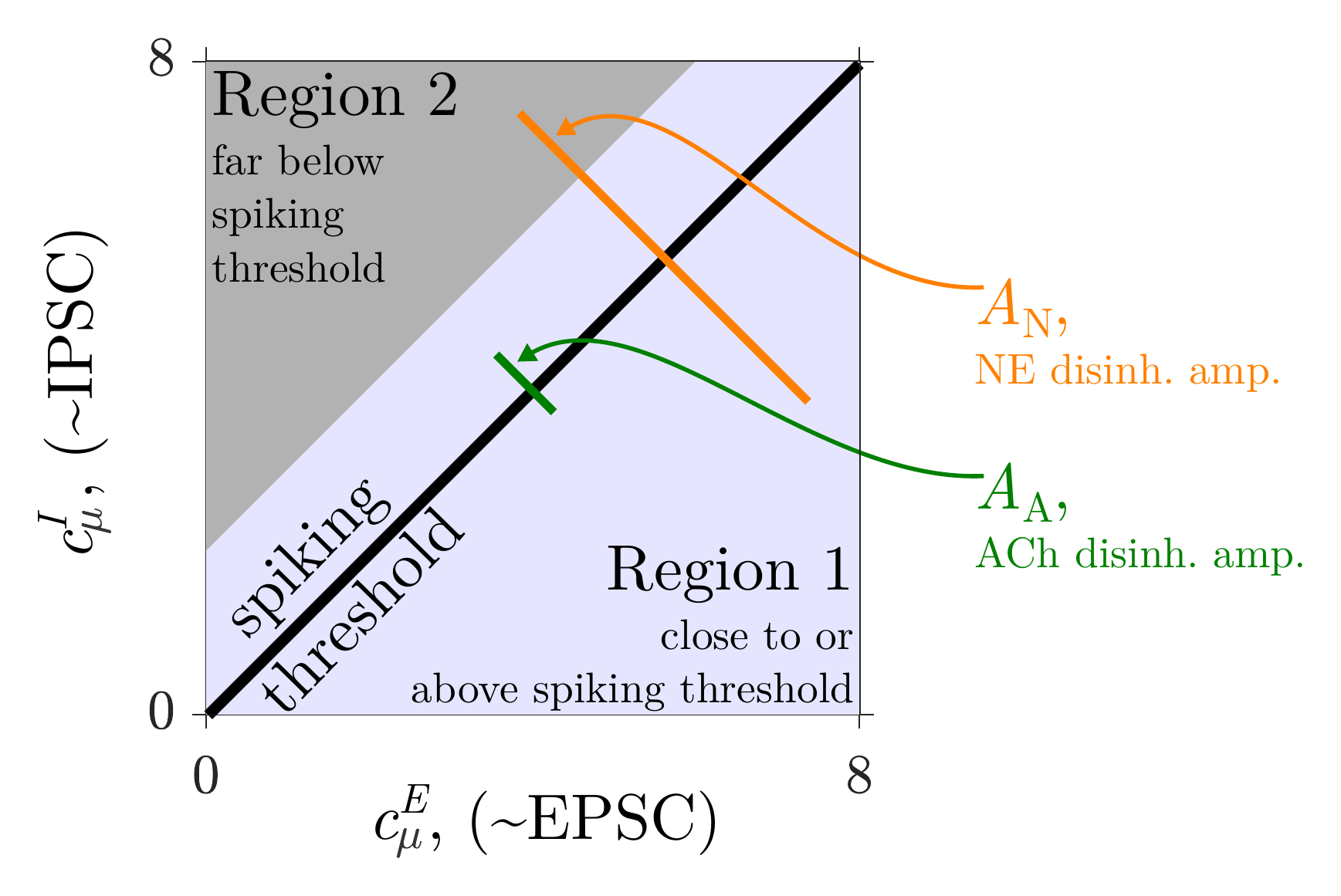}
	\end{center}
	\caption{{\bf Schematic illustration of Regions 1 and 2 in reduced E-I space.} Our derivation of the reduced model relies on the probability of crossing the spiking threshold (black line). When the net current is below threshold (i.e., $d({\bf c}) = c^E_{\mu} - c^I_{\mu} - \theta<0$), a spike could occur as a result of a neuromodulation-dependent disinhibition (characteristic amplitudes of ACh and NE are shown in green and orange, respectively). The plasticity resulting from a spike depends the state of the ACh and NE gates, so averaging over the plasticity mechanisms in our model requires knowing if a threshold crossing is due to pairing with ACh or NE. To make the calculation tractable we separated the reduced E-I space to two regions-- and make a make a different approximation in each region. In Region 1 (light blue, close to or above threshold) we assume that ACh and NE disinhibition are equally likely to cause the neuron to spike, so the contribution of each mechanism is proportional to the pairing rate ($\rho_{\rm A},\,\rho_{\rm N}$). In Region 2 (gray, far from the spiking threshold) we assume that only NE-dependent disinhibition can lead to a spike, so in that region only that neuromodulator is taken into account when computing the average effect on synaptic weights.}\label{fig:schema}
\end{figure}

\newpage

\bibliographystyle{abbrvnat}
\bibliography{PlasticityLearningBibliography}
\end{document}